\documentclass[twocolumn]{aastex631}
\usepackage{hyperref}

\newcommand\mki{Department of Physics and Kavli Institute for Astrophysics and Space Research, Massachusetts Institute of Technology, Cambridge, MA 02139, USA}
\newcommand\eaps{Department of Earth, Atmospheric, and Planetary Sciences, Massachusetts Institute of Technology, 77 Massachusetts Avenue, Cambridge, MA 02139, USA}
\newcommand\cfa{Center for Astrophysics \textbar \ Harvard \& Smithsonian, 60 Garden Street, Cambridge, MA 02138, USA}
\newcommand\gsfc{NASA's Goddard Space Flight Center, Greenbelt, MD 20771, USA}

\newcommand{\tess}{{\it TESS}}
\newcommand{\TESS}{{\it TESS}}
\newcommand{\Kepler}{{\it Kepler}}
\newcommand{\jwst}{{\it JWST}}
\newcommand{\hubble}{{\it HST}}

\newcommand{\rearth}{{\ensuremath{R_{\oplus}}}}
\newcommand{\mearth}{{\ensuremath{M_{\oplus}}}}
\newcommand{\tmag}{{\ensuremath{T_{\rm mag}}}}
\newcommand{\vmag}{{\ensuremath{V_{\rm mag}}}}
\newcommand{\teff}{{\ensuremath{T_{\rm eff}}}}
\newcommand{\tois}{2241}
\newcommand{\accessed}{8 Oct.~2020}

\usepackage{xcolor}
\usepackage{verbatim}
\usepackage{rotating}
\usepackage{longtable}
\usepackage{xfrac}
\usepackage{floatrow}

\usepackage{outlines}
\usepackage{subfigure}

\definecolor{my_color}{HTML}{3a18b1}

\definecolor{new_color}{HTML}{CF0000}
\definecolor{new_black}{HTML}{000000}

\graphicspath{{./}{figures/}}

\accepted{Feb. 17, 2021}

\submitjournal{ApJS}

\shorttitle{TESS Prime Mission TOI Catalog}
\shortauthors{Guerrero et al.}

\begin{document}


\title{The TESS Objects of Interest Catalog from the TESS Prime Mission}


\correspondingauthor{Natalia Guerrero}
\email{nmg@mit.edu}

\author[0000-0002-5169-9427]{Natalia M.~Guerrero}
\affiliation{\mki}


\author[0000-0002-6892-6948]{S.~Seager}
\affiliation{\mki}
\affiliation{\eaps}
\affiliation{Department of Aeronautics and Astronautics, MIT, 77 Massachusetts Avenue, Cambridge, MA 02139, USA}

\author[0000-0003-0918-7484]{Chelsea X.\ Huang}
\altaffiliation{Juan Carlos Torres Fellow}
\affiliation{\mki}

\author[0000-0001-7246-5438]{Andrew Vanderburg}
\altaffiliation{NASA Sagan Fellow}
\affiliation{Department of Astronomy, The University of Wisconsin-Madison, Madison, WI 53706, USA}
\affiliation{Department of Astronomy, The University of Texas at Austin, Austin, TX 78712, USA}


\author[0000-0001-9828-3229]{Aylin Garcia Soto}
\affiliation{Department of Physics and Astronomy, Dartmouth College Hanover, NH 03755}    

\author[0000-0002-4510-2268]{Ismael Mireles}
\affiliation{\mki}

\author[0000-0002-2135-9018]{Katharine Hesse}
\affiliation{\mki}

\author[0000-0003-0241-2757]{William Fong}
\affiliation{\mki}

\author[0000-0002-5322-2315]{Ana Glidden}
\affiliation{\mki}
\affiliation{\eaps}

\author[0000-0002-1836-3120]{Avi~Shporer}
\affiliation{\mki}

% ...

\begin{abstract}
We present \tois\ exoplanet candidates identified with data from the \textit{Transiting Exoplanet Survey Satellite} (\tess) during its two-year prime mission.
We list these candidates in the \tess\ Objects of Interest (TOI) Catalog, which includes both new planet candidates found by \tess\ and previously-known planets recovered by \TESS\ observations. 
We describe the process used to identify TOIs and 
investigate the characteristics of the new planet candidates, and discuss some notable \TESS\ planet discoveries. 
The TOI Catalog includes an unprecedented number of small planet candidates around nearby bright stars, which are well-suited for detailed follow-up observations. 
The \tess\ data products for the Prime Mission (Sectors 1--26), including the TOI Catalog, light curves, full-frame images, and target pixel files, are publicly available on the Mikulski Archive for Space Telescopes.

\end{abstract}




\section{Introduction}
\label{intro}

The Transiting Exoplanet Survey Satellite (\tess; \citealt{ricker2015}) is an MIT-led NASA Astrophysics 
Explorer Mission 
designed to detect transiting exoplanets around the nearest, brightest stars.
During its two-year Prime Mission\footnote{\tess\ launched on April 18, 2018 and the \tess\ Prime Mission ran from July 25, 2018 - July 4, 2020.}, \tess\ observed $\sim70\%$ of the celestial sphere in 26 observing ``sectors,'' resulting in observing times ranging from $\sim$1 month near the ecliptic to $\sim$1 year near the ecliptic poles.

The primary goal of the \tess\ mission is to identify hundreds of small planets ($R_{p} < 4 R_{\Earth}$) and determine the planetary mass for 50 of these small planets with follow-up spectroscopy. Simulations of the Prime Mission \citep{sullivan2015,  Huang2018predictions, barclay2018, barclay2020} predict how many planets \tess\ will detect for a range of periods up to 100 days. 
These simulations predict that \tess\ will find thousands of planets with periods less than 50 days.

\tess\ builds 
on NASA's 
\Kepler\  \citep{borucki2010} and repurposed K2 missions \citep{howell2014}. \Kepler\ and K2 discovered nearly 3,000 confirmed exoplanets with thousands more awaiting confirmation or validation\footnote{NASA Exoplanet Archive, accessed 14 August 2020}.
While the original \Kepler\ mission stared deeply into a single field of 116 square degrees to produce a statistical sample of exoplanets\footnote{\Kepler\ targets were typically 700-1100 pc distant \citep{berger2018}} in the observation ``cone,'' \tess\ is conducting a survey of nearby, bright stars. These targets are the most accessible to both ground and space-based follow-up, necessary for measuring planet mass and density \citep[i.e.][]{huang2018,toi186b}, as well as characterizing atmospheres. Select \tess-discovered planets will likely be good targets for observing with the upcoming James Webb Space Telescope (\jwst) for atmospheric characterization. 
Some \tess\ targets are already being characterized by the Hubble Space Telescope (\hubble), with 12 \hubble\ programs awarded observing time for this purpose to date, e.g. TOI-1231 b \citep{2020hst..prop16181K}, AU Mic b \citep{2019hst..prop15836N,2020hst..prop16164C}, and many others.

By reaching down to small planets transiting bright stars that are suitable for follow-up measurements, TESS is bridging one of the gaps in exoplanet science.
The wide, red bandpass of the \tess\ cameras (600-1000 nm) makes \tess\
capable of detecting Earth-sized and super-Earth-sized exoplanets ($\lesssim 1.75 R_{\Earth}$) transiting M-dwarf stars, which are significantly smaller and cooler than our Sun. M-dwarfs make up the majority (around $75\%$) of stars in our solar neighborhood\footnote{\url{http://www.recons.org/census.posted.htm}} \citep{recons2018, Dole1964}.  For FGK stars, TESS can find planets down to about $R_{p} = 0.8 \rearth$, depending on the planet period and star brightness. 


During the Prime Mission, two basic types of data were collected: 1) small summed image subarrays (``postage stamps'') centered on 20,000 pre-selected targets every two minutes and 2) summed full-frame images (FFIs), measuring $24^{\circ} \times 24^{\circ}$, collected from each of the four \tess\ cameras every thirty minutes. During a typical observing sector, which lasts $\sim$26 days, $\sim$19,000 sets of postage stamps and $\sim$ 1,200 sets of four FFIs are collected.
These data are processed into calibrated light curves by two data processing pipelines described in Section \ref{sec-lightcurves}: the Science Processing Operations Center (SPOC) pipeline for the postage stamps, and the Quick Look Pipeline (QLP) for the FFIs. 
These pipelines identify potential transiting planets by searching for periodic flux decreases, known as Threshold Crossing Events (TCEs), in both the short (two-minute) and long (30-minute) cadence data. 

The \tess\ Science Office (TSO) examines TCEs using the \tess\ light curve and other information 
to identify
planet candidates which would benefit from follow-up observation. 
The light curves are first run through software that eliminates obvious non-planetary signals; the remaining light curves are manually vetted to identify a set of \tess\ Objects of Interest, or TOIs.
 Threshold-crossing events (TCEs) which fall under other categories (such as stellar eclipsing binaries, variable stars, instrument systematics, and non-planet transients) are not included in the TOI Catalog, but are 
 included in the comprehensive TCE Catalog, to be archived at the Mikulski Archive for Space Telescopes (MAST). 
Follow-up observations, both imaging and radial-velocity, are then used to confirm the planet identification and measure planetary parameters.

Over the course of the Prime Mission, 
the SPOC pipeline has produced flux time series (i.e. light curves) for $>200,000$ stellar objects at a two-minute cadence. The QLP has produced light curves for $\sim$16 million stars brighter than \tess-band magnitude, \tmag$=$13.5, observed in the Full-Frame Images (FFIs) at 30-minute cadence.
Additionally, the community produces millions of time series \citep{boumalightcurves, caldwell2020, feinstein2019, tasoc2015, tasoc2017, diamante2020, pathos2019, Oelkers2019} from the FFIs which are not evaluated in this work.

In the Prime Mission, the TSO examined
$55,281$ TCEs from $27,822$ two-minute targets, resulting in $17,296$ candidates for manual vetting. 
The TSO examined $\sim 105,400$ TCEs from $82,490$ unique FFI targets, resulting in $14,629$ candidates for manual vetting.  
Taking into account the overlap in the data sets, the total number of TCEs subject to manual vetting was 32,814 TCEs on 16,489 unique targets.
In total, data collected during the Prime Mission yielded \tois\ TOIs; of the 1676 TOIs not later ruled out as False Positives, 1575 TOIs had periods less than 50 days, and 654 TOIs had radii $R_{p} < 4 \rearth$.

In this work, we first give an overview of \tess's observing strategy (Section~\ref{sec-observing}). Then, we discuss \tess's data collection and processing (Section~\ref{sec-dataprocoverview}). Additional details about the \tess\ instrument, observing strategy, and data processing are described in greater detail in the \tess\ Instrument Handbook \citep{handbook}, data release notes\footnote{\url{https://archive.stsci.edu/tess/tess_drn.html}}, and associated references. We also provide a brief description of light curve generation (Section~\ref{sec-lightcurves}) and the TCE search process (Section~\ref{sec-transitsearch}). The initial stages of the TOI triage and vetting process
are discussed in Section~\ref{sec-TCETriage} and Section~\ref{sec-vetting}. The process by which TOIs are delivered to the community is described in Section~\ref{sec-TOIRelease}. The TOI Catalog is described in Section~\ref{sec-TOICatalog}. Lastly, we turn to a discussion of the \tess\ mission planet candidates found to date, highlighting systems of particular scientific interest (Section~\ref{sec-discussion}). 

\section{Observing Strategy}
\label{sec-observing}
\tess\ is equipped with four wide-field cameras, each with a field-of-view (FOV) of $24^{\circ} \times 24^{\circ}$ and a pixel angular size of $21 ''$ square.  
The four FOVs are arranged to cover a $24^{\circ} \times 96^{\circ}$ region of the sky, the long axis of which is oriented parallel to a line of ecliptic longitude. \tess\ observes each $24^{\circ} \times 96^{\circ}$ field over the course of two orbits (about 27.4 days). Each $\sim27$-day observation period is called a sector.

\begin{figure*}[ht]
    \includegraphics[width=6.5in]{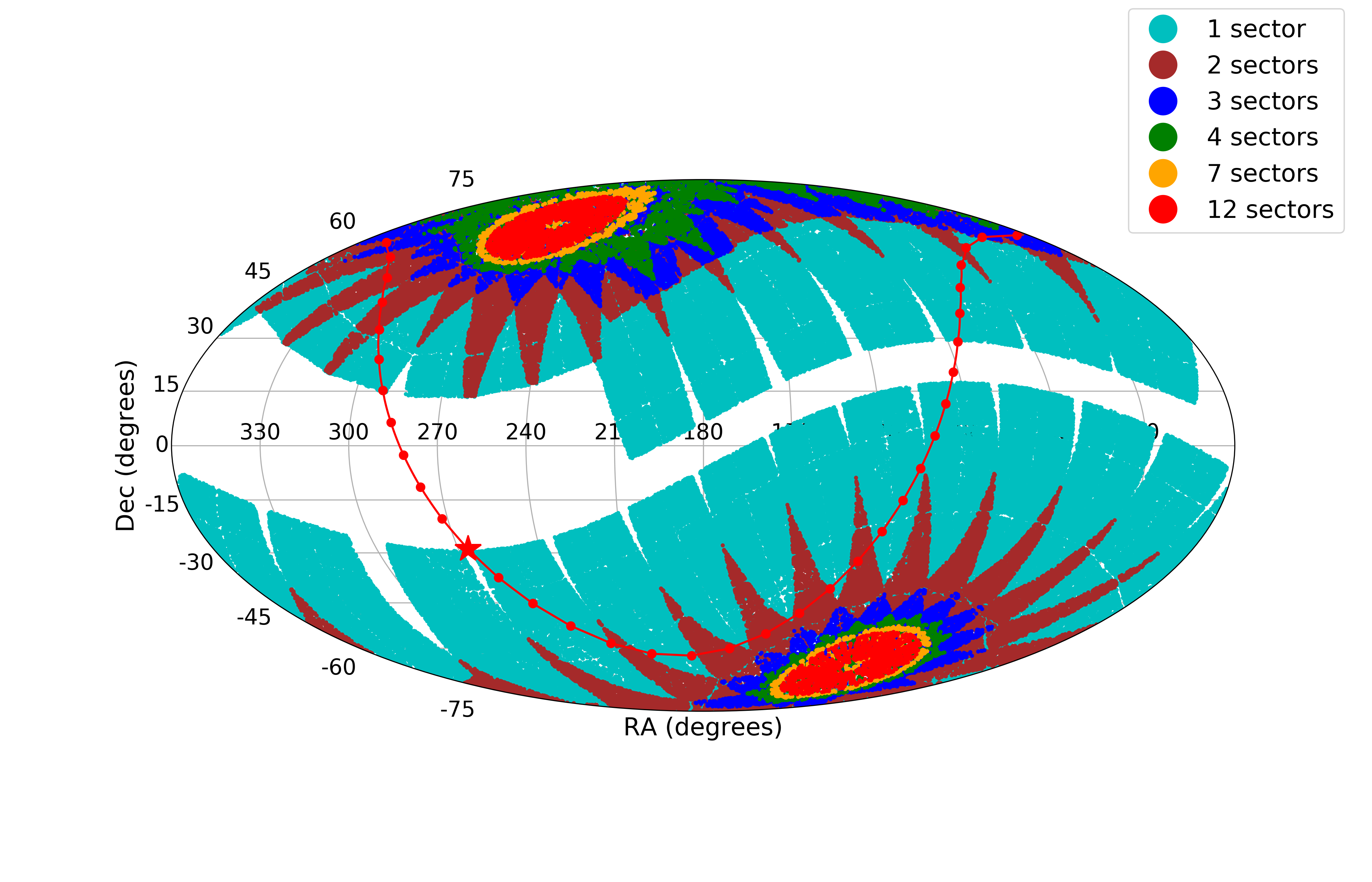}
    \caption{The \tess\ two-year target observation map. Each point in this map represents a target observed at two-minute cadence. The color of the dot represents the number of times it was observed. The U-shaped red curve shows the galactic plane, with the position of the galactic center shown by the red star. Over the course of its two-year primary mission, \tess\ observed 26 sectors for approximately 27 days each, covering $\sim 70\%$ of the sky. Each sector is a $24^{\circ} \times 96^{\circ}$ field of view. 
    The \tess\ ``continuous viewing zones'' ($\sim 351$ days of observation) at each ecliptic pole are clearly visible, as are the regions of the sky observed during more than one sector. In Year 2, during Sectors 14-16 and 24-26, the camera boresight was shifted from the nominal $+54^{\circ}$ orientation northward to an $+85^{\circ}$ ecliptic latitude, due to excessive contamination by stray Earth-light and moonlight in cameras 1 and 2 in those sectors. These gaps will be filled in part during the first Extended Mission.}
    \label{fig:target_coverage}
\end{figure*}

The instrument FOV is generally centered on an ecliptic latitude of $-54^\circ$ (Year 1) or $+54^\circ$ (Year 2), with six sectors in Year 2 centered at $+85^\circ$ (see Figure \ref{fig:target_coverage}).
The ecliptic longitude of the FOV is adjusted by $\sim27^\circ$ from one sector to the next, which results $\sim 27$ days of observation over a large range of ecliptic longitude and an overlap region within $12^\circ$ of the ecliptic poles that is observed for an entire year.

Figure~\ref{fig:target_coverage} shows the two-year sky coverage map for \tess. 
\tess's Prime Mission covered 13 sectors in the Southern Ecliptic Hemisphere during the first year and 13 sectors in the Northern Ecliptic Hemisphere during the second year. The observing period is longest for targets near the ecliptic poles, which were each observed for all 13 sectors.

\tess\ collects an FFI from each camera every 30 minutes over the course of each observation sector. In addition, a set of stars is observed every two minutes in smaller ``postage stamps". A postage stamp typically measures  $11 \times 11$ pixels, but is expanded for brighter stars. These two-minute cadence ``target stars" are chosen from the \tess\ Input Catalog \citep[TIC,][]{stassun2018, stassun2019} based on the Candidate Target Lists (CTLs) contributed by mission stakeholders \citep{handbook}. In order of priority, the categories are 1,920 engineering stars (Photometer Performance Assessment, PPA); all bright ($I<6$) stars; additional potential exoplanet host stars; asteroseismic targets; Guest Investigator targets; and Director's Discretionary Targets\footnote{Full details of how the targets are selected can be found in the instrument handbook (\S 9.4, \citealt{handbook})}. In total, up to 20,000 stars from these categories are observed at two-minute cadence in each 
sector.\footnote{Initially, the target list for two-minute cadence data was limited to the requirement-specified 16,000 target stars. However, after the first three sectors, the Science Operations Center (SOC) team expanded the target list to 20,000 targets beginning with Sector 4, based on tests of the spacecraft compression and pipeline throughput.}

\section{Data Collection and Processing}
\label{sec-dataprocoverview}

\tess\ is in a unique 13.7-day, highly-inclined, highly-elliptical orbit in resonance with the Moon: an observation sector consist of two \tess\ orbits around the Earth. Science data, in the form of 30-minute FFIs and two-minute postage stamps, are collected continuously during each orbit.  At orbit perigee, the data are downlinked from the spacecraft to the Deep Space Network (DSN) and forwarded to the Payload Operations Center (POC) at MIT for decompression and reformatting. 

POC processing reformats the uncompressed data for use by two independent data processing pipelines: the Science Processing Operations Center (SPOC) pipeline and the Quick Look Pipeline (QLP). These pipelines each calibrate the pixel data and use them to produce light curves. In Section~\ref{sec-lightcurves}, we provide a brief overview of the pipelines. 

In a typical sector, \tess\ downlinks roughly 40 GB of compressed target data and 54 GB of compressed FFI data through the DSN.
The postage stamp data products per sector typically expand to 380 GB of target pixel files, 290 GB of collateral pixels (used for calibration), and 18 GB of light curve files and various other data products. The FFI data per sector typically expand to 370 GB of raw FFIs and 710 GB of SPOC-calibrated data. MAST archives the FFI data and target pixel data as well as the ancillary data products SPOC creates. The full collection of full frame images used to search for TOIs from the Prime Mission are available at MAST via\dataset[10.17909/t9-17nt-6c71]{\doi{10.17909/t9-17nt-6c71}} for Sectors 1-13 and\dataset[10.17909/t9-t6cs-8p81]{\doi{10.17909/t9-t6cs-8p81}} for Sectors 14-26. The full collection of two-minute cadence data used to search for TOIs from the Prime Mission are available at MAST via\dataset[10.17909/t9-rm11-zz03]{\doi{10.17909/t9-rm11-zz03}} for Sectors 1-13 and\dataset[10.17909/t9-p87b-kk02]{\doi{10.17909/t9-p87b-kk02}} for Sectors 14-26.

\section{Light Curve Generation}
\label{sec-lightcurves}

The SPOC pipeline converts the postage stamp data into two-minute cadence light curves, while the QLP produces light curves at 30-minute cadence from the FFIs. Both pipelines are described in the following sections. 
Additional details can be found in  \citet{jenkins:KDPH_TPS, Twicken2016},  and Jenkins et al. (in prep.) for the SPOC pipeline, \citet{huanghlsp1,huanghlsp2} for the QLP.

\subsection{SPOC Light Curve Generation}
The SPOC pipeline 
(\citet{jenkins:KDPH_TPS, Twicken2016} and Jenkins et al. (in prep.)) 
processes the two-minute postage stamp data on a per-sector basis. The pipeline calibrates the pixel data; extracts photometry and centroids for each target star; and identifies and removes instrumental signatures. The SPOC pipeline then searches each light curve for TCEs, fits each TCE with a limb-darkened transit light curve model, and performs diagnostic tests to assess the planetary nature of each TCE \citep{Jenkins2016}. The SPOC pipeline also calibrates the FFIs at the pixel level.

\subsubsection{Calibration} 
The SPOC pipeline calibration module (CAL) operates first on both the postage stamp and FFI data to transform the digital counts from the CCDs into flux units \citep[$e^{-} s^{-1}$,][]{KDPH_CAL}.

The sequence of operations is as follows: 1) the module subtracts 2-D fixed pattern noise (2-D black model); 2) estimates and removes the CCD bias voltage (or black level) in each readout row; 3) corrects for non-linearity and gain; 4) measures and removes vertical smear flux due to the shutterless operation of the CCDs, along with the dark current from each column; and finally, 5) corrects for the flat field. Note that the non-linearity, gain, and flat field are pre-flight, on-the-ground measurements. Uncertainties are propagated through each numerical operation and provided with the calibrated pixel values.

\subsubsection{Photometry}
The pipeline module, Compute Optimal Apertures (COA), identifies the optimal aperture for each two-minute target \citep{brysonCoaKADN,bryson:TAD2010SPIE,KDPH_COA}. The module uses the TIC, models for the instrument, such as the Pixel Response Function (PRF), and observations to generate synthetic star scenes for each sector. The star scenes predict the average flux from each star in each pixel of the target star's postage stamp. For sectors 1-13, the pipeline used stellar parameters from TICv7 \citep{stassun2018}. For sectors 14 onward, the pipeline used parameters from TICv8 \citep{stassun2019}. The module analyzes the stellar scene to identify the aperture that maximizes the SNR of the flux measurement. Additionally, the module estimates the contamination due to nearby stars in the apertures, and the fraction of the flux from the target star contained in the optimal aperture.

The aperture size is magnitude-dependent. For bright, saturated targets the photometric apertures can be quite large ($\>$2500 pixels). For stars that are not saturated, typical numbers of optimal aperture pixels in Year 1 range from 20 pixels at \tmag$\,\approx\,$7, to 11 pixels at \tmag$\,\approx\,$10, to 6 pixels at \tmag$\,\approx\,$13. The photometric apertures for target stars with $\tmag < 11$ were somewhat larger in Year 2 to reduce light curve variations due to pointing jitter. 
The crowding in each photometric aperture is estimated from the TIC magnitudes and (proper-motion-corrected) coordinates of nearby stars and a Pixel Response Function that is interpolated at the location of the given target. Flux levels are adjusted for crowding in the pre-search data conditioning (PDC) light curves. The CROWDSAP keyword in the FITS files reports the average fraction of the flux in the photometric aperture that is from the target star.  As CROWDSAP decreases, the photometry becomes less reliable, especially for values $\lesssim$ 0.8. The FLFRCSAP keyword reports the average fraction of the flux of the target star captured in the photometric aperture.

The photometric analysis module (PA) identifies background pixels, then estimates and subtracts a local background from each target star postage stamp.\footnote{This is a new feature of the module for \tess\ as the postage stamps are typically 121 or more pixels (11 by 11), compared to \Kepler's nominal 32 pixels per postage stamp, which were too small to allow for local background estimation.} The module then sums the pixels in the optimal aperture to estimate the brightness of each target star in each frame \citep{twicken:PA2010SPIE,KDPH_PA}. The module also extracts brightness-weighted centroids for each target star in each frame. A subset of 120 bright, unsaturated stars are fitted to the local PRF to extract high-precision centroids that are used to calculate the World Coordinate System (WCS) coefficients for all targets and for the FFIs. Uncertainties are propagated through the numerical operations and reported with the flux and centroid measurements.

The presearch data conditioning module (PDC) \citep{Stumpe2012,Smith2012,Stumpe2014} identifies and corrects for instrumental effects that are highly correlated across the star fields on each CCD. The module first conducts a singular value decomposition (SVD) analysis of the quietest half of the target stars on each CCD to identify the correlated instrumental signatures. The module then fits the most significant eigenvectors to the quiet stars' light curves and formulates empirical prior probability density distributions for the fit coefficients. These distributions provide constraints in a Maximum A Posteriori (MAP) fit for the flux time series for each target star. The module also identifies and records outliers, corrects for the finite flux fraction of the target star's PRF, and corrects for the crowding contamination from nearby stars.  
 The SPOC pipeline propagates uncertainties for the photometry and centroids. The uncertainties for the flux time series are dominated by shot noise from the brightest targets at $\tmag\ \approx 2$, where the ratio of the flux uncertainty to the star’s shot noise is $\sim$1.05, out to approximately \tmag\ = 9, where the flux uncertainty is 25\% above that expected for shot noise from the star. The ratio of the uncertainties to the shot noise rises to 1.78 at \tmag\ =  12, and attains a value of $\sim$4 at \tmag\ = 15. The uncertainties do not account for image motion on timescales less than 2 minutes, or for stellar variability.
 
The uncertainties in the flux-weighted centroids (which are calculated for all stars) vary as a function of magnitude, with typical values of $1\times10^{-4}$ pixels at \tmag\ = 6, $5\times 10^{-4}$ pixels at \tmag\ = 9, and $2.7\times 10^{-3}$ pixels at \tmag\ = 12. The uncertainties for the row centroids of stars brighter than \tmag\ = 6 grow with brightness due to saturation and bleed. The uncertainties do not account for short timescale pointing errors, however, and the typical short timescale scatter at the 2-min level is $\sim5\times 10^{-3}$ pixels after modeling out the pointing history.

The SPOC pipeline has evolved over the course of Year 1 to adapt to changes in the instrument and target list. In Sectors 1-3, the spacecraft jitter impacted aperture photometry, but could be largely corrected using cotrending\footnote{The pointing errors for \tess\ are not stationary and exhibit sporadic excursions on timescales  $\ll$2 min that change the effective point spread function in each two-minute interval, making it difficult to fully correct such behavior on two-minute integrations.}. An on-flight software patch reduced the spacecraft jitter for Sector 4 onward \citep{handbook}. 
However, systematic noise from spacecraft jitter continues to be a source of systematic error removed by PDC, along with rapidly varying scattered light features that occur when the Earth or Moon are within 25$^\circ$ of the boresight of any camera. The updated SPOC codebase, R4.0, processed Year 2 Sectors 20-26 and reprocessed the Year 1 data and Year 2 Sectors 14-19. The update automatically flags scattered light features at the target star/cadence level, which improves the specificity of the data gapping and preserves more data.

\subsection{QLP Light Curve Generation}
The Quick Look Pipeline (QLP; \citet{huanghlsp1,huanghlsp2}) generates light curves from the FFIs, complementary to the SPOC pipeline which produces light curves from the postage stamp data. The QLP separately performs its own calibration of the raw FFIs. For each sector, the QLP produces about half a million light curves. In addition, the QLP always uses all sectors of available data to create light curves (i.e. the QLP runs over multiple sectors of data together). 

\subsubsection{Calibration}
The QLP corrects the raw FFIs for amplifier bias, smear caused by
shutter-less frame-transfer exposure, flat-fields, fixed-pattern noise, and non-linear response.  Amplifier bias and smear corrections are calculated from overscan columns and virtual rows associated with individual FFIs.  Flat-fields, fixed-pattern noise, and non-linearity are corrected using models constructed from pre-launch measurements and on-orbit commissioning data. \citet{vanderspek19} describe these various instrumental effects. The associated calibration procedures and code are included in the \texttt{tica} package and are described more fully in \citep{faus2020}. The calibrated images are in units of number of electrons.

\subsubsection{Photometry}
The QLP uses a circular aperture photometry method to extract light curves for all stars in the TIC with \tess-band magnitudes brighter than 13.5. 
The apertures centers are based on a predetermined astrometric solution derived for each observed frame using stars with \tess\ magnitudes between 8-10. The light curves are extracted using five circular apertures with radii of 1.75, 2.5, 3.0, 3.5, 8.0 pixels. The aforementioned process as well as some of the further steps (see below) are implemented using the various tasks of the FITSH package \citep{fitsh}.

The procedure to produce the target flux time series and background time series are as follows. The photometric reference frame is computed using the median of 40 frames with minimal scattered light. The reference frame provides a high signal-to-noise image to compute the difference images, using a direct subtraction of the photometric reference frame from the observed frames. 
The final flux time series of the star is calculated as the sum of the constant flux of the star and the variability of the star over time. The constant flux is taken from the theoretical calculation in the TIC and the zero-point magnitude measured for \tess. This method effectively de-blends the light curve from contamination by an additional star inside the aperture. We measure the variability over time in the difference images. 
The flux measured in the aperture in the difference images is a sum of the stellar variability and the background variability.
To calculate the background variability (from uniform faint background sources), we draw annuli around the target star in the difference images. The annuli are drawn to have approximately the same area as the aperture. The background variability at a given time point is calculated as the median of the background pixels in the annulus in the difference image after iterative outlier rejection.
 We estimate the background time series independently as a sum of some constant background flux and background variability over time. The constant term is calculated by subtracting the star flux (from the TIC catalog) from the total flux in the aperture. We assume the background per pixel is locally uniform in the vicinity of the target star, its aperture, and background annuli; therefore, the constant background per pixel in the aperture should be the same as in the annulus. The background variability is calculated as the difference between the background in the reference image annulus and in the difference image annulus.

The QLP then selects one of the five light curves extracted from the five different apertures. Because the QLP's primary goal is to detect transiting planets (as opposed to preserving stellar variability), the QLP selects the optimal aperture by minimizing photometric scatter in high-pass-filtered light curves. An optimal aperture for stars in each magnitude range (13 linear bins between \tess\ magnitudes of 6-13.5) is then selected by determining which aperture size produces the smallest photometric scatter in the magnitude bins.
The photometric precision roughly follows the prediction from the \citet{sullivan2015} estimation for the majority of the stars and has a lower noise floor\footnote{The on-orbit performance of the TESS photometers is better than pre-flight calculations from \citet{sullivan2015}, which has been traced to an underestimate in their assumed telescope aperture} (approximately 20 ppm) for the brightest stars when the spacecraft operates nominally.
\TESS\ light curves usually contain low-frequency variability from stellar activity or instrumental noise, which must be filtered before the small, short-duration signals caused by transiting planets can be readily detected. Following \citet{vj14} and \citet{shallue18}, the QLP removes this variability by fitting a B-spline to the light curve and dividing the light curve by the best-fit spline. Outlier points caused by spacecraft momentum dumps or other instrumental anomalies are masked out prior to detrending. To avoid distorting any transits present, the spline is iteratively fit, $3\sigma$ outliers are removed, and the spline is refit while interpolating over these outliers \citep[see Fig. 3 in][]{vj14}. 

\section{Transit Search}
\label{sec-transitsearch}

To search for transits, 
both the SPOC and QLP pipelines
phase-fold processed light curves with a
large number of trial periods to search for repeating transit-like drops in brightness. Any drop in brightness that passes a specified threshold requirement is called a threshold-crossing event or TCE. Each TCE has an associated Data Validation (DV) Report that provides a detrended and phase-folded light curve in addition to auxiliary data products.

\subsection{SPOC Transit Search}
For each sector, the SPOC pipeline conducts a search for transiting planet signatures in the systematic error-corrected light curves from PDC\footnote{\texttt{PDCSAP\_FLUX} column in the light curve file} using a module called Transiting Planet Search (TPS). Targets with transit-like features that exceed the threshold for statistical significance of 7.1$\,\sigma$, have two or more transit signals, and that pass a number of other statistical vetoes \citep{seader:vetoes2013ApJS,Seader2015,Twicken2016} are then processed by the Data Validation module (DV). The module runs a suite of tests on each transit signature to help inform the classification of the events. The SPOC pipeline also conducts periodic multi-sector searches, allowing for the detection of much-longer-period transiting planet signatures as described in Section~\ref{subsec-spocdv}.

\subsubsection{Transiting Planet Search Algorithm}

The Transiting Planet Search algorithm (TPS) searches for signatures of transiting planets by adaptively characterizing the power spectral density of the observation noise, then estimating the likelihood of a transit over a range of trial transit durations and orbital periods as a function of time over the observations, quantified as the Single Event Statistic \citep[SES;][]{jenkins2002, jenkins:TPS2010,jenkins:KDPH_TPS}.

TPS folds the SES time series over a range of trial orbital periods to identify the strongest peak in the folded SES time series and designates TCEs in the light curves with three requirements: a Multiple Event Statistic (MES) in excess of 7.1, a passing grade on a series of transit consistency tests, and at least two transits\footnote{The \Kepler\ Science Pipeline transit search required a minimum of three transits to declare a TCE \citep{jenkins:KDPH_TPS}}. The MES represents the transit SNR multiplied by the correlation coefficient between the true signal waveform and the trial waveform best matched over the grid-search parameter space (duration, period, epoch). 

TPS also provides an estimate of the effective noise ``seen'' by a transit of a given duration, called the combined differential photometric precision \citep[CDPP --][]{christiansen2012,jenkins:KDPH_TPS}. The CDPP can be used to infer a lower limit for transit depth detectable at a given duration and number of transits.

\subsubsection{Data Validation}
\label{subsec-spocdv}

The Data Validation (DV) module fits a limb-darkened transit model to the light curve for each target with a TCE and employs the same whitening filter as is used by TPS to account for non-stationary correlation structure in the observation noise \citep{Li2019}. 
A number of diagnostic tests are conducted to inform the vetting process \citep{Twicken2018}. 

\begin{figure*}[ht]
    \includegraphics[width=7.5in]{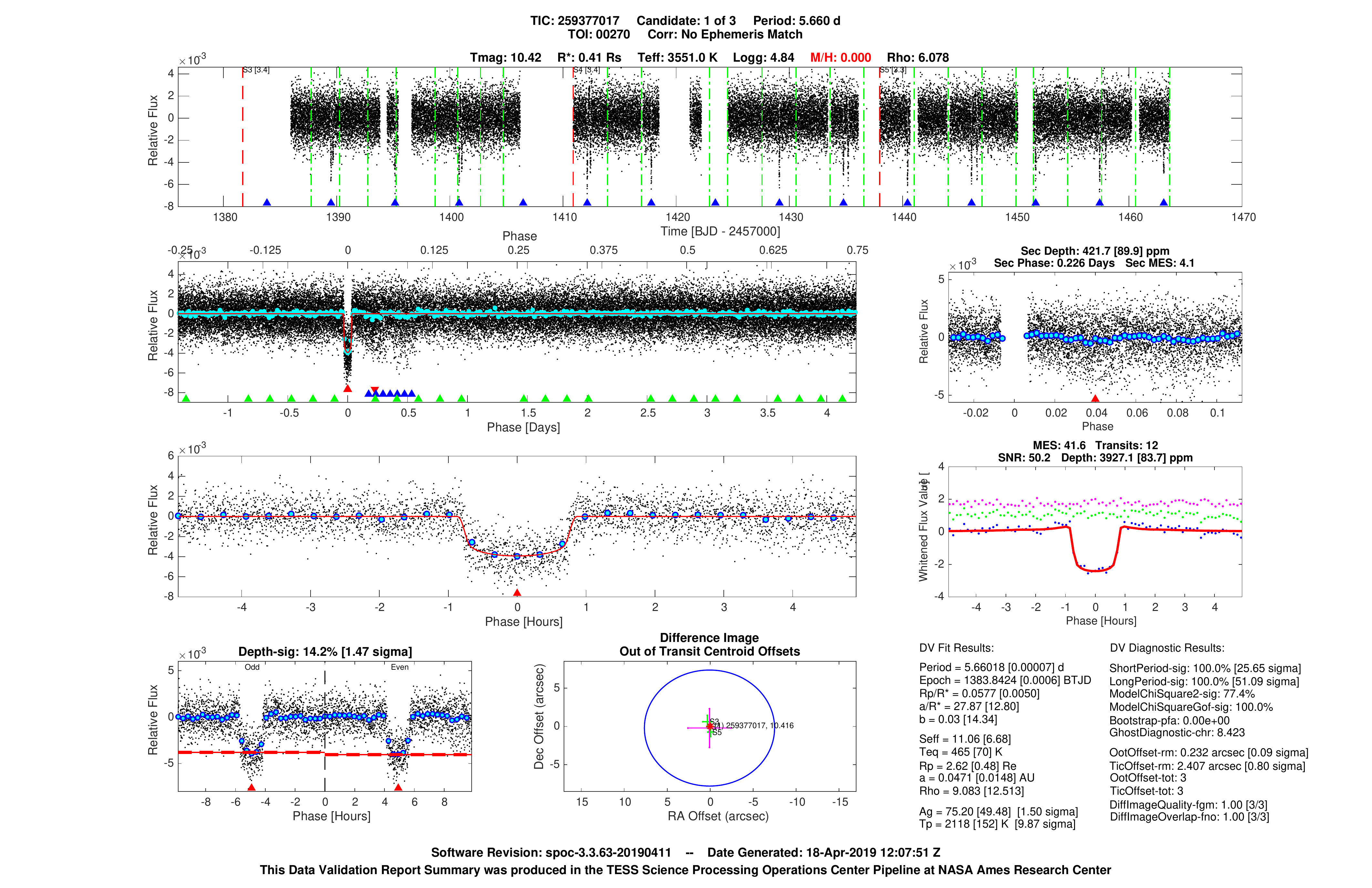}
	\caption{SPOC Data Validation (DV) summary report plots for TOI 270 b.  
	\textit{Top row:} The detrended \TESS\ light curve with transits marked by blue triangles. Sector boundaries are marked with red vertical dashed lines. Momentum dumps (described in Section~ \ref{subsec-falsepos}) are marked with green vertical dashed lines. \textit{Second row left:} The \TESS\ light curve folded on the planet candidate's orbital period. The black points are all data points, cyan points show data binned at $\sfrac{1}{5}$ the fitted transit duration, and the red line is a model fit to the transit. \textit{Second row right:} The best candidate for a secondary transit found in the phase-folded light curve.  \textit{Third row left:} Zoom-in on the phase-folded light curve. \textit{Third row right:} The whitened transit fit is indicated by the red line; the binned residuals from the fit are displayed in green, and the out-of-phase residuals from the fit are shown in magenta. \textit{Fourth row left:} A comparison of the phase-folded odd transits with the phase-folded even transits. The red dashed lines indicate transit depths. The title line reports the significance of the difference between the odd and even transits. \textit{Fourth row center:} The centroid offset plot showing the right ascension and declination offsets of a transit source with respect to the position of the target star (as determined by the out-of-transit image centroid). The centroid offsets for each sector are shown with green crosses; the mean centroid offset over all sectors is shown with a magenta cross. The location of the target star is marked by a red star. The 3$\sigma$ radius of uncertainty for the transit source offset is shown by a blue circle. \textit{Fourth row right:} Fitted planet parameters and data validation diagnostic test results. After its initial detection and vetting using these data products, a three-planet system around TOI 270 was eventually validated by \citet{guenthertoi270}.}
    \label{fig:spocdv}
\end{figure*}

The DV module produces a one-page PDF summary report  for each TCE, a PDF full report, and a PDF mini report for each target. The full report for each target with TCEs includes the results of all the tests along with diagnostic figures and run time warnings and errors. The mini-report for each target with TCEs combines the one-page summary along with the most informative diagnostic tests and graphics from the full report. An example one-page summary is shown in Figure~\ref{fig:spocdv}. The DV module also produces a FITS file of the time series it analyzes as well as those presented in the diagnostic plots, and an XML file with numerical model fit and diagnostic test results \citep{SDPDD2018}.

The DV module generates several metrics used in
the assessment of the TCEs to help eliminate background eclipsing binaries and background scattered light features.

Difference images for each sector that represent the difference between the out-of-transit fluxes adjacent to each putative transit and in-transit fluxes are used to  
obtain the difference image centroids to identify the location of the source of the transit-like features. These fit results are used to inform the difference image centroid offset statistics, which are averaged over multiple sectors where applicable and indicate whether the source of the transit-like feature is consistent with the location of the target star. 

The difference image centroid offsets are computed separately using two approaches for target reference location: (1) the mean-proper-motion-corrected catalog location of the target star during transit in the given sector, and (2) the PRF-based centroid of the mean out-of-transit image in the given sector. The first method is robust against crowding (because the difference image features only the source of the transit signature), and has been observed to be more reliable than the second for crowded TESS images. The second method is not robust against crowding, but is more accurate in uncrowded fields because the centroid bias common to the difference image and out-of-transit image is removed.
 
Uncertainties in the centroid offsets are formally propagated from uncertainties in the calibrated pixels from which the centroid offsets are computed. In addition, an error term is added in quadrature to the propagated uncertainties in the centroid offsets to account for systematic effects in the centroid offsets. 
This was done in the Kepler Science Pipeline as well. For Kepler, the plate scale was smaller (4 arcsec/pixel) and most targets were observed in 17 quarters, permitting significant averaging of the quarterly centroid offsets; 
the quadrature error term was set so that the difference image centroid offsets were not significant at the 3-$\sigma$ level if they were $<$0.2 arcsec. For TESS, the plate scale is larger (21 arcsec/pixel) and a majority of targets are only observed in a single sector; 
the quadrature error term is set so that the centroid offsets are not significant at the 3-$\sigma$ level if they are $<$ 7.5 arcsec.

The ghost diagnostic is formulated as the ratio of the detection statistic of the flux time series obtained with the optimal photometric aperture to the detection statistic of the flux time series obtained from a halo about each target's optimal aperture. 

After conducting the diagnostic tests (useful, e.g., for detecting contamination of the background by light scattered from nearby bright variable stars), the DV module masks out the transits of the current TCE \footnote{One full transit duration prior to and following each transit are masked along with the transits in the search for additional TCEs in the SPOC pipeline.} and calls TPS to search for additional TCEs in the light curve (up to a maximum of six TCEs per target star). DV also compares the orbital periods between TCEs associated with a given target star to flag cases where the periods are statistically identical; this can indicate that the TCEs are primary and secondary events in stellar eclipsing binaries rather than transiting planet signals. 

TPS and DV are also run on light curves for stars observed in multiple sectors, increasing sensitivity to longer-period and smaller planets. Multi-sector planet searches were run for Sectors 1-2, Sectors 1-3, Sectors 1-6, Sectors 1-9, and for Sectors 1-13; in Year 2, for Sectors 14-16, 14-19, 14-23, and 14-26. In these multi-sector runs, TCEs could appear in each of the sectors searched for the multi-sector range, or in a subset. In SPOC/TPS, targets are processed in later multi-sector runs only if they were observed in sectors not included in a previous multi-sector run. For example, a target observed only in Sectors 1 and 2 would not have been included in the Sector 1-3 run.

Following processing, the SPOC pipeline delivers the TCEs and data validation products to the POC. POC conducts a final review and quality assessment, then prepares the products for submission to MAST\footnote{\url{https://archive.stsci.edu/tess/bulk_downloads.html}}. MAST ingests these data, along with the FFIs, postage stamp data, light curves, and cotrending basis vectors \citep{SDPDD2018}. The public data products are accompanied by POC-produced data release notes for each sector. For certain sectors before March 2019, MAST released the FFIs ahead of the TCEs and data validation products. This staggered release, which efficiently released early \tess\ data to the community, allowed observers to follow up target stars in the given sector while still visible for the given season.

\subsection{QLP Transit Search}
The QLP carries out a corresponding but independent analysis to the SPOC pipeline. Complementary to SPOC pipeline, the QLP also aims to identify TCEs along with diagnostic information. 

\subsubsection{Box-Least Squares Transiting Planet Search}

The QLP begins by stitching light curves together from all available sectors for each target. The QLP then searches the light curves for periodic transits using the Box Least Squares algorithm \citep[BLS;][]{bls}. The search is performed for periods ranging from $0.1$ days to half the length of the longest baseline expected for the given target star. The number and spacing of frequencies searched by BLS is adapted to the total baseline in the light curves, similar to \citet{vanderburg16}.

Any signal with a signal-to-pink-noise ratio \citep[SNR, as defined by][]{vartools} $>9$ and BLS peak significance $>9$ is designated as a TCE. The BLS peak significance is defined as the height of the BLS peak in the spectrum compared to the noise floor of the BLS spectrum.

To search for additional transiting planets in a light curve with a detected signal, the QLP clips out transits with the period of the strongest BLS peak, masking one full transit duration prior to and following each transit along with the transit. QLP re-runs the search until no further significant peaks are found. 

\subsubsection{QLP Data Validation}

\begin{figure*}[ht]
    \includegraphics[width=7.5in]{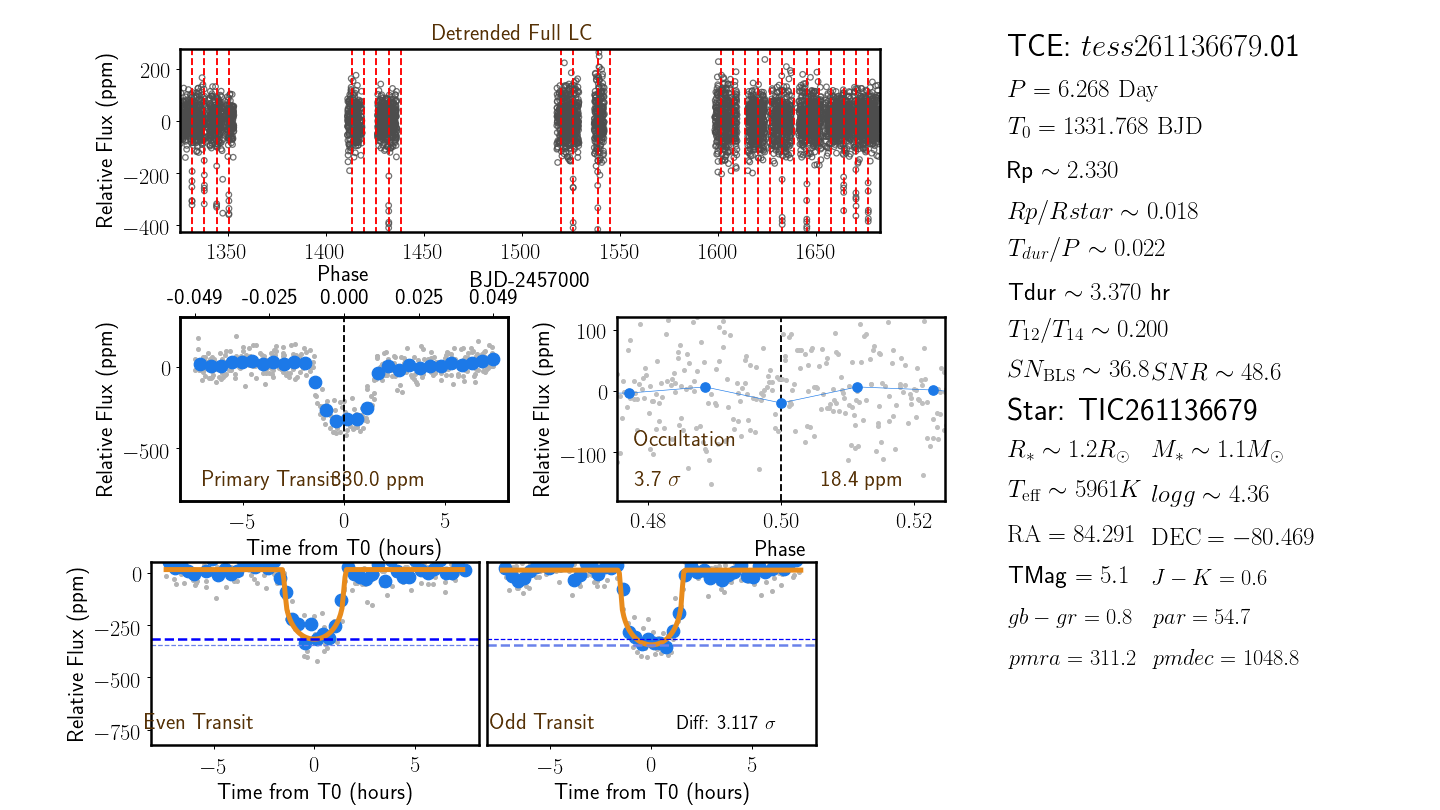}
	\caption{QLP Data Validation (DV) summary report plots for $\pi$ Men c \citep{huang2018}. \textit{Top row:} The detrended full light curve with all available sectors stitched together. Transits are marked by red vertical lines. \textit{Second row left:} Phase-folded light curve. The grey points are all data points and the blue points are binned relative to the transit window size to guide the eye. \textit{Second row right:} The phase-folded light curve at phase 0.5 for a secondary eclipse test. \textit{Third row:} Even and odd transits. The model fit is shown in orange. The depths of even transits are marked by darker dashed lines, and odd transits by lighter dashed lines. A significant difference in the depths ($> 5{\sigma}$) would suggest that the TCE is an eclipsing binary at twice the period. 
	\textit{Right-hand column:} Planet and stellar parameters. }
    \label{fig:qlpdv}
\end{figure*}

The QLP data validation report provides diagnostic plots to aid in TCE classification. The first page, shown in Figure~\ref{fig:qlpdv}, is a summary page. 

Additional pages in the full QLP data validation report provide additional metrics for decision-making. These plots are useful for diagnosing whether the source of transit-like variability is on target or from a nearby blended source, which is particularly important for FFI targets. For example, the QLP compares the depth of the transit signal measured in multiple apertures of different size. If the depth of the transit increases with aperture size, that is usually a sign that the variation is from a nearby source and not from the target. The QLP also searches within $155 ''$ of the target star for other light curves with transits which match in period and epoch. The QLP reports also check whether the epoch and period of the TCE matches with known planets or false positives \citep{Coughlin2016, EBLM,KELTFP, WASPFP}. 

\section{TCE Triage}
\label{sec-TCETriage}

Most of the signals detected by the SPOC/TPS and QLP pipelines are not due to transiting planets, but instead are caused by astrophysical phenomena such as stellar variability, eclipsing binary stars, or non-astronomical instrumental artifacts. 
We use a ``triage'' process to identify transit-like events in the TCEs produced by both the SPOC and QLP pipelines. The goal of triage is to separate out transit-like events from stellar variability and instrument noise.

We identify the signals most likely to be planetary transits using a combination of automatic classification algorithms and visual inspection. We identify the best planet candidates in a two-step process: a partially automated triage in which the least-likely TCEs to be planet candidates are efficiently removed followed by a more labor-intensive team vetting process. 

\subsection{Triage of SPOC TCEs}
TESS-ExoClass (TEC)\footnote{\url{https://github.com/christopherburke/TESS-ExoClass}} 
provides an automated method for the vetting team at the TSO to identify potential planet candidates from the postage stamp TCEs generated by the SPOC pipeline. TEC borrows heavily from the successful automated planet candidate vetting tools from the \Kepler\ Robovetter \citep{Mullally2015,Coughlin2016,Thompson2018}. 

Here we summarize the TEC process. We are actively improving TEC and describe here its latest form, which was largely stable in the latter half of the first year of \tess\ observations (Sectors 6-13), and had few changes in Year  2 of the Prime Mission. TEC changed rapidly during the first four \tess\ sectors as the tests and their thresholds were tuned in order to match the manual classification effort performed by the TSO. TEC python code is open source and classification tables from TEC for observed \tess\ sectors are available. However, due to the changing nature of the thresholds, detections from Sectors 1-6 are less uniformly selected relative to the latter sectors where TEC thresholds and algorithms were more stable.

TEC establishes an initial filter that removes TCEs likely due to instrumental systematics.  The initial filter criteria ensure that the signal continues to be significant when an alternative detrending is applied to the flux time series and individual transit events have the shape expected for a limb-darkened transiting planet. The TCEs that pass this initial triage step are subject to additional tests. These tests identify false positives due to stellar binaries, astrophysical variability, or instrument systematics \citep{Brown2003,Coughlin2016,Mullally2018}. TEC also flags TCEs with large centroid offsets and those occurring at the same time as a spacecraft momentum dump \citep{Li2019}. 

Remaining TCEs are classified into three Tiers. Tier One corresponds to TCEs that passed all the above criteria and did not receive any warning flags. The Tier One TCEs represent the highest quality and most likely detections to result in planet candidates. Any TCE which receives at least one warning flag in the above tests is placed in the Tier Two category. However, TCEs with a warning flag for a significant secondary eclipse or a warning flag for coherent out-of-transit variability are assigned to the Tier Three category.  The Tier Three category is  predominantly made up of stellar eclipsing binary detections. An exception to prevent a candidate from inclusion in Tier Three is made if the secondary eclipse could be consistent with reflected light from a planet rather than a self-luminous body. Only the Tier One and Two lists are sent through to the team vetting stage.

As an example of TEC processing results, for the original 1458 TCEs generated by the SPOC pipeline for Sector 12, 540 TCEs passed the initial TEC triage stage. Of those 540, 45 TCEs were assigned to the high quality Tier One list; 315 TCEs were assigned to the Tier Two list; and 180 were assigned to Tier Three. 

\subsection{Triage of QLP TCEs}
\label{subsec-qlptriage}
TCEs generated by the QLP number in the tens of thousands per sector, as compared to typically 1500 TCEs per sector from the SPOC pipeline. An aggressive triage process is therefore needed to cull these signals to a few hundred high-quality potential planet candidates to be reviewed by eye. We use Astronet-Triage \citep{yu2019},  a version of the AstroNet convolutional neural network (CNN) classifier \citep{shallue18} modified to work on \TESS\ data, to separate eclipse-like events from instrument systematics and stellar variability. \citet{yu2019} uniformly labeled QLP TCEs (i.e., careful manual vetting of an inclusive sample of TCEs) from sectors 1-5 to use as input training data for the new AstroNet-Triage CNN.

For each sector, Astronet-Triage ingests tens of thousands of TCEs brighter than \tmag $=$ 13.5 and selects several thousand (3000-4000) targets as potential planet candidates.

Next, this number is further reduced by setting a brightness threshold of \tmag $=$ 10.5, prioritizing target stars most accessible to ground-based observers. In addition, the TCEs with estimated planet radii greater than 35 $\rearth$ are removed. For TCEs with unknown stellar radius (and therefore unknown planetary radius), a transit depth cut is implemented at 6$\%$ for stars hotter than 3500~K or stars cooler than 4000 K with a proper motion of less than 10 milliarcseconds/year.
In sectors 1-11, stellar parameters for imposing these cuts (and for the QLP DV reports) came from Gaia DR2 \citep{gaiamission, gaiadr2}. If Gaia DR2 data were unavailable, the most recent TIC catalog was used. For sectors 1-11, this was TICv7; for Sector 12 onward, the pipeline referred to TICv8 for stellar parameters. Following the release of TICv8, QLP reprocessed the light curves for the Year 1 TOIs using the updated stellar parameters.

Following these cuts, only a few hundred TCEs remain for visual inspection. The QLP generates DV reports for these targets, including centroid information, difference image comparison,  multi-aperture comparison, and an MCMC fit to the folded transit \citep{emcee}. Also, at this stage, the QLP flags candidates which match in period and epoch against catalogs of known eclipsing binary false positives \citep{collins2018, Triaud2017}.

\section{Vetting}
\label{sec-vetting}
\begin{figure*}[ht]
    \includegraphics[width=7in]{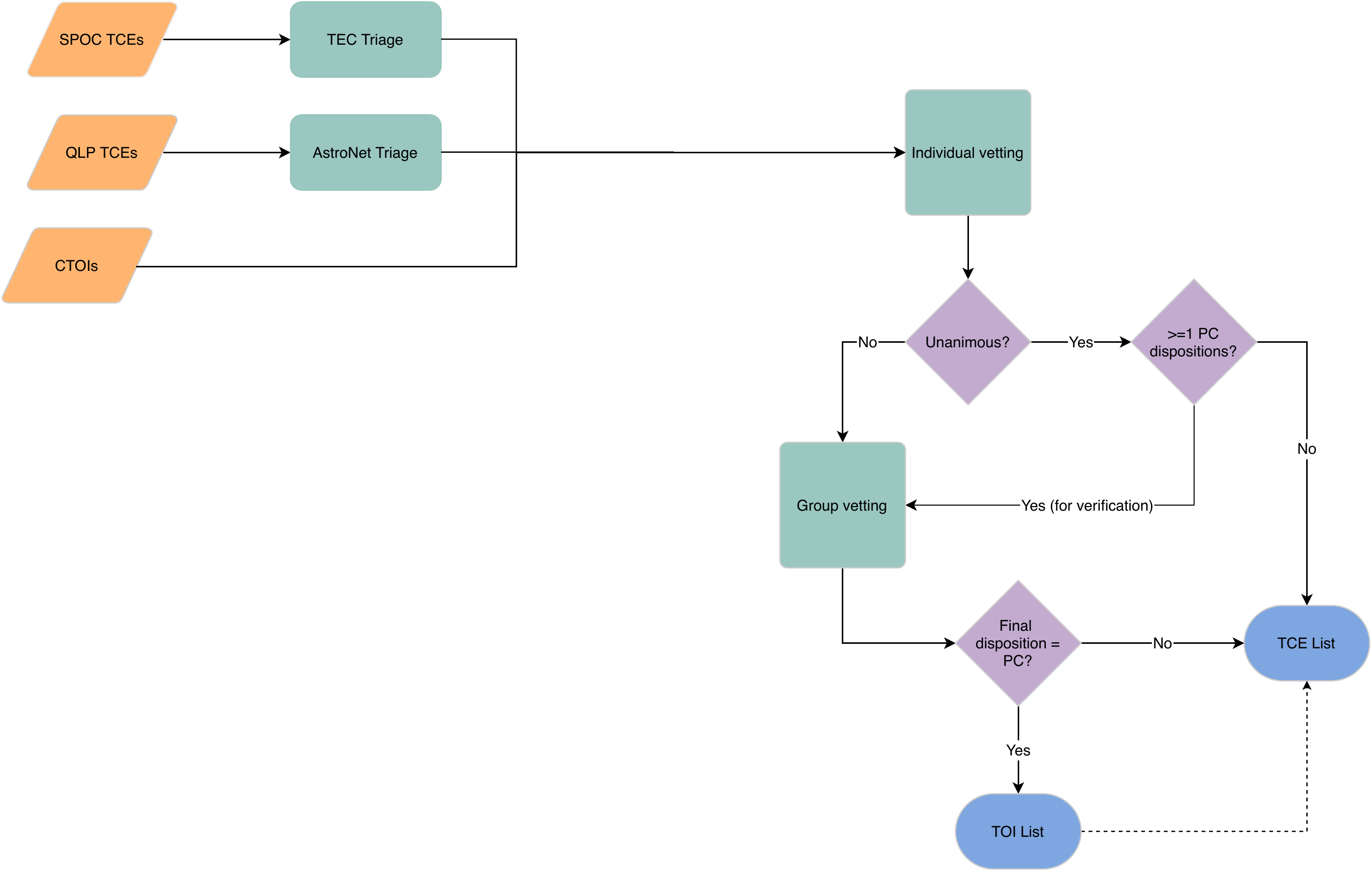}
	\caption{Flow chart of the TOI triage and vetting process. Orange blocks are sources of TCEs. Green blocks are process steps. Purple diamonds are decision points. Blue blocks are outcome products. The TOI process triages TCEs from the pipeline of record and the community, then vets any TCE with at least one Planet Candidate disposition from individual vetting. The TOI process produces the TOI List, which is a subset of the list of vetted TCEs.}
    \label{fig:vettingflow}
\end{figure*}

Vetting refers to visual inspection of DV products at the MIT TSO to determine which TCEs will become part of the TOI Catalog. Group vetting is the main vetting process, while individual vetting is used to reduce the number of surviving planet candidates for cases where computer algorithm triage has passed a large number of candidates. The vetting process was ``rehearsed" using K2 C17 data before \tess\ science operations began \citep{crossfield2018}. Individual vetting was heavily used in sectors 1-5 before TEC and AstroNet-Triage were available. The vetters are a rotating team of about 40 volunteers from the TSO and wider exoplanet community at various levels of expertise. New vetters are trained extensively by the TSO. Figure~\ref{fig:vettingflow} describes the flow of TCEs through the vetting process. In addition to inspecting TCEs from the two pipelines, the vetters examine candidates identified by the community (CTOIs, as explained in Section \ref{subsec-ctois}). The goal of vetting is to provide planet candidates to the community but is not intended to be 100\% complete.

\subsection{Individual Vetting}
\label{subsec-iv}

Individual vetting is visual inspection of TCE data products for classification by a team of vetters working independently. Individual vetting is used if a large number of TCEs make it past triage, with the goal of reducing the number of TCEs that must be inspected during the more definitive and labor-intensive group vetting process. In individual vetting, the vetters use custom software known as the \tess\ Exoplanet Vetter (TEV) to disposition TCEs. Dispositions are based on the DV products, including the raw light curve, phase-folded detrended light curve, odd-even plots, difference image and out of transit images, finder charts of nearby stars, and estimated planet radius based on reported stellar parameters \citep{Twicken2018}.
TEV also provides vetters with a text box to provide comments for Group Vetting to review. 

Each target is vetted by at least three and no more than five vetters. To ensure all potential planet candidates are considered, any TCE vetted as a planet candidate (PC) or undecided (U) by any vetter is sent through to Group Vetting, where it will be carefully reviewed by a small group with more expertise. Individual vetters are trained to be generous in their dispositioning so that edge cases are preserved and sent to Group Vetting. TCEs vetted unanimously as not a planet candidate or undecided are assigned a disposition of eclipsing binary (EB), stellar variability (V), or instrument noise/systematic (IS) based on the majority rule and exit the vetting process. Table \ref{tab:dispositions} lists all disposition labels.
In cases of a tie between non-planetary dispositions, eclipsing binary is used. TCEs that leave the vetting process as EB, V, or IS are not included in the final TOI Catalog.

\begin{deluxetable}{cl}
\tablecaption{Definitions for Vetting Dispositions.
\label{tab:dispositions}}
\tablecolumns{2}
\tabletypesize{\scriptsize}
\tablehead{
\colhead{Disposition} & \colhead{Definition}
} 
\startdata
PC & Planet Candidate; U-shaped transit \\ 
EB & Eclipsing Binary; V-shaped transit \\
V & Stellar Variability; sinusoidal transit \\
IS & Instrument Noise or Systematic; often ramp- or transit-shaped \\
U & Undecided; ambiguously transit-like \\
\enddata
\tablecomments{
The vetting process uses features of the transit light curve, as well as other observables, to determine the disposition of each TCE.
}
\end{deluxetable}

Early in the mission, before TCE triage was largely automated by TEC and AstroNet-Triage, individual vetting was part of the main process for identifying TOIs. By Sector 6, however, automatic triage had significantly reduced the need for individual vetting.  Individual vetting remains useful for teaching vetters the characteristic appearance of \tess\ data and training them to make accurate classifications.

\subsection{Group Vetting}
\label{subsec-groupvetting}
The TCEs that pass through the triage process (and individual vetting, if applicable) proceed to a Group Vetting session where a team of at least three vetters visually inspect each TCE. Each sector typically yields about 200 TCEs for Group Vetting. The TSO vetting teams of experienced experts meet in up to four sessions per week, handling about 50 TCEs in two hours per session. Group vetters include persons with expertise on historical \Kepler\ data analysis, and new vetters who are brought on board through extensive training, starting with participation in individual vetting. 

The main goal of Group Vetting is to promote suitable TCEs to the TOI list and to remove false positives from consideration. Vetting relies on a team of vetters who can visually discern subtleties in the data better than computers and can impose physicality and consistency priors. Vetting by eye also enables us to keep track of performance differences between the data pipelines. It is possible that Group Vetting will be partially automated in the future, but the TSO has focused on enhancing the automated triage stage so far.

Group Vetting reviews key metrics in the DV reports that can expose false positives. For example, off-target, transit-like signals may be distinguished from on-target planets by examining the location of the centroid in and out of transit. Movement of the centroid in the difference images to a star 1 pixel away or more shows that the pipeline triggered on a TCE that is likely caused by a background eclipsing binary. Off-target TCEs can also be revealed by comparing the depth of transits in light curves extracted from differently sized apertures on the QLP DV reports, or using the ghost diagnostic flag in the SPOC/TPS DV reports. 
If the transit depth increases with aperture size, the TCE signal usually comes from a nearby source  and can be removed from consideration for the TOI Catalog.  

During Group Vetting, supplementary information is considered such as parameters from Gaia DR2 (which were implemented by SPOC in Sector 14-26), and information from ExoFOP-TESS\footnote{ExoFOP-TESS, or the Exoplanet Follow-up Observing Program, is the website which supports community follow-up of TESS targets by organizing community-gathered follow-up data and derived astrophysical parameters, \url{https://exofop.ipac.caltech.edu/tess/}} about known planets and/or previous observations.

In addition to setting the disposition of TCEs, Group Vetting assigns follow-up priority levels for each of the five subgroups (SGs) of the \tess\ Follow-Up Observing Program (TFOP). Planet candidates with $R_p < 4 \rearth$ are \tess's highest priority, and those orbiting brighter stars are more likely than those orbiting fainter stars to be amenable to mass measurements with precise radial velocity instruments. To assign initial priorities, we consider the promise of each system for precise radial velocity (PRV) observations as well as the possibility that false positives remain that cannot be excluded by \tess\ data alone. We calculate the radial velocity (RV) semi-amplitude using the \citet{chen:2017} planetary mass-radius relation, and then estimate the telescope time required to measure a 5-$\sigma$ mass using a 4-m class precise RV facility such as HARPS. 

The most promising TOIs (typically with \vmag\ $\lesssim$ 11) are assigned priority 1 for photometry (SG1) and reconnaissance spectroscopy (SG2) because it is necessary to know if they are false positives or not before investing in precise radial velocity follow-up. TOIs with $SNR < 11$ are assigned priority 2. Priority 2 targets often orbit fainter stars and have smaller expected RV signals.
Larger planet candidates ($R_p > 4 R_\oplus$), or small candidates orbiting stars which are likely too faint for RV follow-up are given priority 3 for SG1 and SG2. 

The other SGs (high-resolution imaging, precise radial velocities, and space-based photometry) have priorities initially set to 4 (meaning that observations are on hold), and priorities are updated on ExoFOP-TESS following results from SG1 and SG2. Known planets, discovered previously in surveys such as WASP, HAT, and KELT, are priority 5 for all follow-up groups (no further observations needed). Finally, we note that even low-priority planet candidates get observed by SG1 and SG2, and that many of the follow-up steps proceed in parallel because of the practical constraints imposed by telescope schedules and the scientific interests of those carrying out observations.

\subsection{False Positives}
\label{subsec-falsepos}

We find false positives in the TCE light curves during vetting and triage. These false positive TCEs are not made into TOIs. However, vetting and triage cannot catch all false positives: we expect follow-up observations will find some fraction of TOIs to be false positives. In the Prime Mission, $565/\tois$ TOIs (about $25\%$) have been identified as False Positives from follow-up observations. We note which TOIs are false positives with a note in the ``Public comment'' field of the catalog, and lower the TFOP priority for those candidates.

Though a full description of the characteristics of \tess\ false positives is beyond the scope of this paper, we list some common features of false positives here. Some light curve features indicative of non-planet signatures are specific to the \tess\ instrument and spacecraft.

For example, spacecraft momentum dumps may create signals that mimic transit-like events in depth and duration. These momentum dumps occur when the reaction wheels on the spacecraft are periodically de-spun to relieve momentum built up from solar radiation pressure imbalance on the spacecraft. Momentum dumps briefly affect the spacecraft pointing for 1-2 frames. In the beginning of Year 1, there wer 3-4 momentum dumps  per orbit. In Year 2, momentum dumps were reduced to 1-2 times per orbit. For early sectors, the SPOC pipeline and the QLP were not able to distinguish transit-like events due to momentum dumps from actual transits. Visually, there is a slight distinction, in a sharp ingress and a slow egress different from a planet transit with ingress and egress having similar slopes. Vetters used the transit shape and the known occurrence times to eliminate events suspected to be caused by momentum dumps.

To aid vetting, the SPOC pipeline added momentum dump markers to the detrended light curves in the DV summary reports, starting with Sector 3. The QLP used the quaternion information (three-dimensional spacecraft orientation, velocity, and rotation) from the spacecraft to mask frames in which momentum dumps were occurring, starting with Sector 1. The SPOC pipeline masks data on cadences identified by the POC as directly associated with momentum dumps in PDCSAP light curves. Such cadences are identified in SPOC pipeline archive products with a dedicated reaction wheel desaturation event data quality bit \citep{SDPDD2018}.

In addition to spacecraft momentum dumps, the pipelines flagged and removed cadences affected by scattered light
\citep{handbook}. A full detailing of these features, in addition to occasional thermal and power events disrupting data acquisition, can be found in the \tess\ Data Release Notes \footnote{\url{https://archive.stsci.edu/tess/tess\_drn.html}}.

Vetters can identify false positive TCEs by a few tell-tale signs: a visible secondary eclipse which cannot physically be planetary in nature; a mismatch between odd and even transit depths caused by a stellar binary at twice the indicated period; a shift of the centroid to another nearby star; and increasing transit depth with aperture size, indicative that a signal is from an object nearby on the sky and not the suspected target star.
 
TCEs are especially prone to being false positives as \tess\ pixels are large enough to include multiple stars. Blended stellar signals from a single pixel can lead to confusion between on-target planetary transits with nearby eclipsing binaries and some types of stellar variability. For the majority of cases, we can recognize the contribution to a star at a minimum distance of 1 pixel away.

Some TCEs have only very subtle differences in transit depth between odd and even transits or weak secondary eclipses that require careful review via group vetting. 
Cases that cannot be resolved on TESS data alone are passed on to TFOP as TOIs for additional observations and detective work with the goal of improving reliability without compromising completeness. The follow-up efforts are described in depth in Section~\ref{subsec-tfop}.

We rely on follow-up observations to identify false positives which were not caught in triage or vetting.
It is very difficult, if not impossible, to identify the star that is the source of the transit signal from the \tess\ data alone when there are two (or more) stars within the same TESS pixel. 
In these cases, where the transit signal could be planetary if on either star, we release the target as a TOI and follow-up observations (by subgroup 1) are used to identify the star showing the transit signal. 
Reconnaissance spectroscopy (obtained by TFOP subgroup 2) is used to confirm the parameters of the primary star and identify TOIs which are stellar eclipsing binaries where both stars show lines in the spectrum (spectroscopic binary type 2). In addition, low-precision RVs, obtained from a few spectra, are used to identify TOIs showing large RV variation of a few 10 km/s, indicating that the transiting object is too massive to be a planet (spectroscopic binary type 1). The other common source of TOI False Positives is our improved understanding of host star parameters. The inclusion of Gaia DR2 data in TICv8 gave a more accurate determination of stellar radius compared to TICv7. For some stars, the new TICv8 stellar radius is bigger, which changes the transiting object radius to be much too large to be a planet.

\subsection{Numbers and Timeline}
Large numbers of TCEs are delivered per sector from each pipeline. The SPOC pipeline produces,
on average,
about 1,200 TCEs per sector, out of about 20,000 two-minute cadence targets observed. Before triage, QLP produces 50,000 TCEs per sector (with a \tmag $=$ 10.5 cutoff for vetting), from about half a million FFI light curves. Each sector yields about 100 total TOIs from the SPOC pipeline and the QLP. Of the 1,250 TOIs released during Year 1, 761 were from the postage stamps and 489 were from non-postage-stamp regions of the FFIs. 651 TOIs had data from a single sector and 599 had data from two or more sectors. Of the 991 TOIs released during Year 2, 494 were from the postage stamps and 497 were from the non-postage-stamp regions of the FFIs. 274 TOIs from Year 2 had data from a single sector and 717 had data from two or more sectors.

Of the total \tois\ TOIs from the full Prime Mission, 256 TOIs were known planets and 1,985 were new planet candidates. We detected 251 of the 1,543 known planets ($\sim16\%$) around $\tmag < 13.5$ host stars originally detected via the transit method. 
We expect this fraction to be small, as many of these known planets were \Kepler\ discoveries with the SNR of transits of many of these planets to be too low for \tess.
For the 864 known planets detected via other methods and orbiting host stars with $\tmag < 13.5$, we detected 5 as new TOIs. We expect the number of radial velocity planets \tess\ re-observed to be low because of transit probabilities being low in general for RV-detected planets. The known planet TOIs are discussed in more depth in Section~\ref{subsec-kps}.

Given the different detrending and detection methods employed by the two different pipelines, we do not expect the two pipelines to always detect the same signals in the same data. Of the 230 TOIs SPOC detected in Sectors 6-13\footnote{Because Sectors 1-5 did not use AstroNet triage, these sectors were not considered in this comparison.} with $\tmag < 10.5$ host stars, the QLP BLS detected 206. 
The QLP AstroNet triage (see Section \ref{subsec-qlptriage}) excluded 58 of these in automatic triage and visual triage rejected an additional 29 TCEs, 
giving a total of 119 TOIs detected by both SPOC and QLP. 
Future upgrades to AstroNet will aim to improve its recall of small and low S/N planet candidates detected by SPOC. 
In sectors 14-26, of the 146 TOIs SPOC detected, QLP BLS detected 137. 
The QLP AstroNet triage excluded 35 of these in automatic triage, and visual triage rejected an additional 15, 
giving a total of 87 TOIs detected by both pipelines.

TESS aims to survey nearby, bright stars rather than conduct an exhaustive census. When one signal is detected by one pipeline but not the other, we rely on the human vetters to determine if the transit detection is valid using the vetting criteria described in Sections \ref{subsec-iv} and \ref{subsec-groupvetting}.

\TESS\ data are processed rapidly once they have been downloaded from the spacecraft. Once the raw FFIs arrive at the TSO, processing by the QLP is completed within about 7-10 days, yielding a large set of TCEs for triage. Concurrently, the SPOC pipeline  produces calibrated FFIs and DV products for two-minute cadence TCEs within 7-14 days from receipt of data, including a formal review process. TEC triages the SPOC pipeline data to produce a reduced set of TCEs in 1-2 additional days. Vetting takes 1-2 weeks per sector and an additional 2-3 days to prepare the TOIs for release. Overall, \tess\ data and TOIs generally reach the public via MAST about a month after the end of a sector. 

\section{Releasing TOIs to the Community}
\label{sec-TOIRelease}
Following group vetting, and a final check for accuracy, the TOIs are released to the community on a variety of platforms. 

\subsection{TOI Delivery}

The TOI Catalog includes planet candidates suitable for follow-up observations, as well as a smaller number of previously-known exoplanets. TOIs are released on a public-facing webpage at \url{https://tess.mit.edu/toi-releases/} and updated automatically (twice daily) on ExoFOP-TESS. The first TOI Catalog was delivered to MAST on December 2, 2018 \footnote{\url{http://archive.stsci.edu/tess/bulk_downloads/bulk_downloads_tce.html}}. MAST archives a static version of the TOI Catalog on a monthly basis. The living TOI Catalog, updated every sector and intermittently between sectors, is described by this paper.

Targets are assigned consecutive TOI IDs. Multi-planet systems are assigned suffixes (.01, .02, etc.) mirroring the suffixes assigned for the TCE. TEV automatically generates a comma-separated values (CSV) file with all necessary parameters for each TOI from the vetted TOI list. Each TOI comes with a table of parameters, a DV summary page, and a full DV report.

Some TOIs may be held back from delivery if, for example, the target requires reanalysis (for instance, to better constrain or confirm its orbital period or to identify/disentangle additional candidates in the light curve). Additionally, if a target shows a marginally-detected transit in a single sector of data, but is expected to be in the \tess\ field of view in a following sector, the team may opt to wait to see if the transit becomes more pronounced before releasing it. Even though SPOC/TPS and QLP/BLS are not designed to search for single-transit events, they are occasionally detected when the transit happens to line up with either spurious noise events or to a data gap in the phase-folded light curves examined by TPS/BLS. We do include these single transit candidates in the TOI catalog, with the orbital period parameter removed before release. If there is another transit of the candidate in a later sector, we update the TOI in the catalog with the correct period. 

\subsection{TOI Updates}
The online TOI Catalog at tev.mit.edu and ExoFOP-TESS is continuously updated to provide the most recent information to the community. We update entries in the TOI catalog in certain cases, including: 

\begin{enumerate}
\item TOIs which appear in several sectors accumulate more observations, which improve the determinations of period, epoch, and other parameters. The TOI catalog is therefore updated with parameters from the longest baseline of data for each TOI, with a preference for SPOC postage stamp data, largely due to its two-minute cadence and review process as the pipeline of record. 

\item TOIs on the ExoFOP-TESS platform have updated TFOP Working Group priorities as each working group makes progress with follow-up observations. 

\item When TFOP identifies TOIs as false positives, their initial disposition in the TOI catalog remains unchanged (to preserve the candidate's history), but a separate TFOP working group disposition is updated on ExoFOP-TESS.~The TFOP working group disposition is updated in the TOI catalog as well.
\end{enumerate}

\subsection{Retired ``Alerts" Process}
The TSO used an ``Alerts" process to get candidates to the community as quickly as possible during Sectors 1-3.  
The goal of the Alerts was to release TOIs for timely follow-up before the full data sets from the sector were publicly available on MAST. Alerts were posted on the MIT TEV Alerts Portal and on MAST as the first high-level science product of the mission. All the TESS-DATA-ALERTS products that were made public are available at MAST via \dataset[https://doi.org/10.17909/t9-wx1n-aw08]{https://doi.org/10.17909/t9-wx1n-aw08}. 
Following Sector 3, the FFI data were processed and released in advance of the completion of the TOI process. 

\subsection{Community TOIs}
\label{subsec-ctois}

Community TOIs (CTOIs) are objects of interest identified by pipelines in the community and reported on ExoFOP-TESS, often in bulk uploads not tied to the official data release from any particular sector. ExoFOP emphasizes CTOIs should only be reported if there is no TOI with the same period. 
The TOI team periodically checks for matches on TIC ID and period  between the CTOI list hosted on ExoFOP and the comprehensive collection of TCEs considered for vetting on the TEV platform.

169 CTOIs from the Prime Mission are now TOIs; 106 of these were recovered from TCEs mistakenly ruled out in pre-vetting triage \footnote{ExoFOP-TESS, accessed \accessed}.
All CTOIs which have been promoted to TOIs were also TCEs in the SPOC pipeline or the QLP, and recovered from those pipelines. For example, two CTOIs from the Planet Hunters \tess\ project \footnote{\url{http://www.planethunters.org}} were promoted to CTOIs and then released as TOIs.

In the Prime Mission, $\sim40\%$ of the 841 CTOIs reported on ExoFOP matched in period and epoch to TCEs with reports on TEV found by the TSO, meaning TCEs which passed triage and were considered for vetting. For the remaining $\sim60\%$ of CTOIs, which did not match with TCEs on TEV, about half corresponded to TCEs from either the QLP or SPOC analysis, and the other half did not have data products from either pipeline.
For the CTOIs which did not match in ephemeris to TCEs on TEV, but which had TCE data products from either SPOC or QLP, many were excluded from vetting in the standard pre-vetting triage process. 
For example, TCEs from the QLP below the $\tmag = 10.5$ threshold were not considered for vetting; other TCEs were ruled out as non-planetary transit events in pre-vetting triage. Single-transit events reported as CTOIs and which were also SPOC TCEs were sometimes mistakenly classified in triage as EB events (described in more detail in Section \ref{subsec:single-transits}). 
These rejected TCEs are re-considered in the CTOI review process. 
The vetting team may promote a CTOI to a TOI if the CTOI identifies a quality planet candidate from the SPOC or QLP pipeline mistakenly ruled out in the triage or vetting process.
CTOIs which do not match with TCEs on TEV, and which also do not have data products from either the QLP or SPOC pipeline, could still be valid events, but were not found in the initial run of either pipeline, and would need to be recovered manually in a reprocessing step.

\subsection{Naming \tess\ Planets}
\label{subsec:tessname}
The project will assign \tess\ planet names to TOIs which are confirmed planets, either with a mass measurement or validated by another method. The \tess\ planet name will have the same major number as the TOI ID. Planet suffixes (b, c, etc.) will follow the suffixes used for publication of the planet, under the general convention to assign first in order of validation, and then period. Said another way, if a planet is discovered with a shorter period than planet b in a given system, the new planet will get the next consecutive letter as a suffix. Planet names will be assigned by the project after a TOI has been confirmed and appears on the NASA Exoplanet Archive. \tess\ planet name will also appear as an additional  column in the TOI Catalog and propagated to MAST and ExoFOP-TESS. TOIs which are known planets from other surveys and which meet the same requirement as TOIs of mass measurement or other validation will get a \tess\ planet name. Assigning \tess\ Planet names to known planets will highlight \tess's ability to recover these planets. Although the \tess\ Prime Mission is focused on measuring sizes and masses for small exoplanets, the broader exoplanet community reaps the benefits of a diverse catalog of \tess\ planets which have been validated without a measured mass.

\section{TOI Catalog Description}
\label{sec-TOICatalog}

The TOI Catalog is a living collection of the \tess\ mission planet candidates from the Prime Mission and beyond. The list includes both \tess-discovered planet candidates (denoted PC) and known planets (denoted KP) from past surveys. The living catalog is available online \footnote{\url{https://tess.mit.edu/toi-releases/}}. This same living list is the version of the catalog posted to MAST and ingested to ExoFOP-TESS to be augmented with information from follow-up observations. Archived versions of the catalog are available on MAST\footnote{\url{https://archive.stsci.edu/missions/tess/catalogs/toi/previous/}}.

\subsection{Catalog Overview}

A TOI is a target initially considered by the project to be a planet candidate based on the TESS light curve and other information available at the time it is inspected by the TSO. 
As TOIs are followed up by additional observations, the project tracks the TFOP Working Group disposition for each target in the catalog, and whether the target is a known planet from a previous survey (KP), promoted to a confirmed planet (CP), retracted as a false alarm (FA), or demoted as a false positive (FP). The CP category includes planets with confirmed masses and statistically validated planets that do not yet have measured masses.

The columns in the TOI catalog uniquely identify each candidate, its provenance, and, in the case of the comments column, particular issues or notable features of a TOI. 
The TOI team also provides to the community a ``TOI+" table with a more expansive set of parameter columns useful for planning follow-up observations, including stellar log \emph{g}, stellar radius, planet radius, planet equilibrium temperature, stellar effective temperature, signal-to-noise, centroid offset flag (``TRUE'' when TOI had a centroid offset greater than 5 sigma), and TFOP priorities. 

We provide a sample of the TOI catalog with stellar and planetary parameters subdivided for ease of reading in Table~\ref{tab:toicatalog_1} and Table~\ref{tab:toicatalog_2}, respectively.  The TOI catalog columns (headers in bold-face type) are as follows:
\begin{itemize}
    \item {\bf Source Pipeline} Name of pipeline which provided the TOI period, epoch, transit duration, and transit depth. Either ``QLP" or ``SPOC.''
    \item {\bf Stellar Catalog} Version of TIC or other catalog (e.g. Gaia) which sources the provided stellar parameters
    \item {\bf TIC} Unique identifier from the TESS Input Catalog.
    \item {\bf Full TOI ID} Unique identifier for each TOI planet candidate for the star. 
    \item {\bf Signal ID} Identifier for each planet candidate in a system; ordered by time of release as a TOI.
    \item {\bf \tess\ name} \tess\ planet name for confirmed planets (see Section \ref{subsec:tessname})
    \item {\bf TOI Disposition} ``PC" for planet candidate and ``KP" for known planet from a previous survey
    \item {\bf EXOFOP Disposition} Provided by ExoFOP-TESS. TFOP Working Group disposition following additional observation (CP for confirmed planet, PC for planet candidate, FP for false positive, FA for false alarm, and KP for known planet). Updated on TEV each time new TOIs are released. 
    \item {\bf TIC Right Ascension } In degrees. All coordinates in equinox J2000.0, and epoch 2000.0. 
    \item {\bf TIC Declination} In degrees. All coordinates in equinox J2000.0, and epoch 2000.0. 
    \item {\bf \tmag} In magnitude. Includes uncertainty.
    \item {\bf \vmag} In magnitude. Includes uncertainty.
    \item {\bf Epoch} Barycentric-corrected epoch of first transit in TESS Julian Days. $\text{BTJD}=\text{BJD}-2457000$ \footnote{\citet{SDPDD2018}}. Includes uncertainty.
    \item {\bf Orbital Period} In days. Can be blank, as is the case for single transits. Includes uncertainty.
    \item {\bf Transit Duration} In hours. Transit duration from first to fourth contact. Uncertainty column can be blank if duration is an estimate.
    \item {\bf Transit Depth} In ppm. Uncertainty column can be blank if transit depth is an estimate.
    \item {\bf Sectors} Space-separated string of sectors contributing data to TOI parameters.
    \item {\bf Public Comment} From TOI group vetting, including whether follow-up is in progress, if the event is a single transit, and the catalog name of the planet if it is previously known. Can be blank. 
    \item {\bf Alerted} Date TOI was released for the first time.
    \item {\bf Edited} Date of most recent update to TOI parameters.
\end{itemize}

The TOI list is provided in the AAS Journals machine-readable format and also as a CSV file with header rows detailing the contents and creation of the list.

\begin{verbatim}
   #TESS Objects of Interest
   # File Created: 2020-10-18
   # Sectors: S0001-S0026
   # Date of previous TOI list release: 2020-08-12
   Source Pipeline,Stellar Catalog,TIC,
   Full TOI ID, Signal ID, TESS Name, TOI Disposition,
   EXOFOP Disposition,TIC Right Ascension,
   TIC Declination,TMag Value,TMag Uncertainty,
   VMag Value,VMag Uncertainty,
   Orbital Epoch Value,	Orbital Epoch Error,
   Orbital Period Value,Orbital Period Error,
   Transit Duration Value,Transit 
   Duration Error,
   Transit Depth Value,Transit Depth Error,
   Sectors,Public Comment,Alerted,Edited
\end{verbatim}

Usually, all of these fields are populated in the TOI catalog with the exception of period and period uncertainty (for single transits), the \tess\ name field, and the comments field. 

\movetabledown=0.5in
\begin{rotatetable*}
\begin{deluxetable*}{llllcccclllllll}
\tablecaption{Example selection from the TOI catalog (Part1)
\label{tab:toicatalog_1}}
\tablecolumns{20}
\tabletypesize{\scriptsize}
\tablehead{
\colhead{Source} & \colhead{Stellar} & 
\colhead{TIC} & \colhead{Full} & 
\colhead{Signal} & 
\colhead{TESS}
& \colhead{TOI} &  \colhead{EXOFOP} & 
\colhead{TIC RA} & \colhead{TIC Dec} & 
\colhead{\tmag} & \colhead{\tmag\ Unc.} & 
\colhead{\vmag} & \colhead{\vmag\ Unc.} & 
\colhead{} \\
\colhead{Pipeline} & \colhead{Catalog} & 
\colhead{} & \colhead{TOI ID} & \colhead{ID} &
\colhead{Name} & \colhead{Disp.} & \colhead{Disp.} & 
\colhead{(deg)} & \colhead{(deg)} & 
\colhead{(mag)} & \colhead{(mag)} & 
\colhead{(mag)} & \colhead{(mag)} & 
\colhead{} 
} 
\startdata
spoc &TICv7 & 281459670 & 110.01 & 1 & 110 b & KP &KP& 5.618692 & -59.942551 & 11.632 & 0.018 & 12.283&0.057 &...  \\ 
spoc &TICv7 & 355703913 & 111.01 & 1 &  111 b & KP &KP &0.774485 & -62.469342 & 13.129 & 0.018 & 13.857& 0.137 &...\\ 
spoc &TICv7 & 388104525 & 112.01 & 1 &  112 b & KP & KP&55.93344 & -65.193856 & 11.547 & 0.018 & 12.314 & 0.057 &...\\ 
spoc &TICv7 & 97409519 & 113.01 & 1 &  113 b & KP & KP&332.714323 & -30.749674 & 12.143 & 0.018 & 12.724  & 0.092 &...\\ 
spoc &TICv7 & 25155310 & 114.01 & 1 &  114 b & KP &KP& 63.37389 & -69.226789 & 10.555 & 0.018 &10.994  & 0.012 &...\\ 
spoc &TICv7 & 281541555 & 115.01 & 1 &  115 b & KP &KP & 6.702439 & -56.316124 & 12.98 & 0.018 &13.783 & 0.126 &...\\ 
spoc &TICv7 & 238176110 & 116.01 & 1 &  116 b &KP &KP& 357.84526 & -70.152863 & 10.991 & 0.019 &11.98 & 0.022 & ...\\ 
qlp & Gaia DR2 & 322307342 & 117.01 & 1 &  117 b& PC &KP& 15.005896 & -58.90478 & 11.734 & 0.019 &12.312 & 0.046&...\\ 
spoc & TICv7 &266980320 & 118.01 & 1 & 118 b& PC &CP& 349.556843 & -56.903885 & 9.154 & 0.017 & 9.809 & 0.003 &...\\ 
spoc &TICv7 &278683844 & 119.01 & 1 &  & PC & & 99.237983 & -58.015237 & 9.234 & 0.018 &10.07 & 0.003 &...\\ 
spoc &TICv7 &278683844 & 119.02 & 1 &  & PC & & 99.237983 & -58.015237 & 9.234 & 0.018 &10.07 & 0.003 &...\\ 
\enddata
\tablecomments{
The TOI catalog table lists the parameters of each TOI and its host star. This table (Part I) gathers together the stellar parameters selected from the TOI catalog. The planet parameters continue on the table on the following page.
}
\end{deluxetable*}
\end{rotatetable*}

\movetabledown=2.5in
\begin{rotatetable*}
\begin{deluxetable*}{lllllllllllll}
\tablecaption{Example selection from the TOI catalog (Part 2) \label{tab:toicatalog_2}}
\tablecolumns{19}
\tabletypesize{\scriptsize}
\tablehead{
\colhead{}&
\colhead{Epoch} & \colhead{Epoch} &  
\colhead{Orbital Period} & \colhead{Orbital Period} & 
\colhead{Transit} & \colhead{Transit Dur.} & 
\colhead{Transit Depth} & \colhead{Transit Depth} & 
\colhead{Sectors} & \colhead{Public} & 
\colhead{Alerted} & \colhead{Edited}  \\
\colhead{} &
\colhead{Val. (TJD)} & \colhead{Err. (TJD)} &  
\colhead{Val. (days)} & \colhead{Err. (days)} & 
\colhead{Duration (hrs)} & \colhead{Err. (hrs)} & 
\colhead{(ppm)} & \colhead{Err. (ppm)} & 
\colhead{} & \colhead{Comment} &
\colhead{} & \colhead{} 
} 
\startdata
...& 1328.040464 & 0.000446 & 3.174305 & 4.40E-05 & 2.723339 & 0.030983 & 15603.6931 & 151.40271 & 1,2 &HATS-30 b &  2018-10-22  & 2020-08-03 \\ 
...&1326.106369 & 0.001084 & 2.106098 & 6.90E-05 & 1.574766 & 0.083967 & 12430.50297 & 528.39557 & 1,2 &HATS-34 b &2018-10-22 & 2020-08-03 \\ 
...&1327.40960 & 0.000119 & 2.499804 & 2.47E-06 & 2.885454 & 0.0141 & 15075.418 & 53.067 & 1,2,3,4,7,11 &WASP-119 b&2019-05-07 & 2020-08-03 \\ 
...&1327.053085 & 0.000623 & 3.372877 & 0.000147 & 2.63426 & 0.043644 & 17163.60419 & 212.40587 & 1 & WASP-124 b &2018-09-05 & 2020-08-03 \\ 
...&1327.520787 & 0.000121 & 3.288783 & 2.30E-06 & 3.408833 & 0.014136 & 7061.865079 & 20.379316 & 1,2,...,13 &WASP-126 b &2019-05-07 & 2020-08-03 \\ 
...&1329.90069 & 0.001812 & 4.742745 & 0.000262 & 2.408768 & 0.137056 & 13420.03179 & 517.9781 & 1,2 &HATS-46 b &2018-10-22 & 2020-08-03 \\ 
...&1326.689274 & 0.00026 & 2.798594 & 4.90E-05 & 2.366459 & 0.022664 & 16879.44984 & 101.11358 & 1 & WASP-91 b &2018-09-05 &2020-08-03 \\ 
...&1328.4809 & 0.00133 & 3.5855 & 0.00038 & 3.173 & 0.123 & 6130 & 3.88166 & 1,2 & HATS-68 b &2018-09-05 & 2020-08-03 \\ 
...&1329.199631 & 0.001235 & 6.036061 & 0.000637 & 2.123832 & 0.09961 & 1704.034261 & 56.202583 & 1 &HD 219666 b&2018-09-05 & 2020-08-03 \\ 
...&1327.61510 & 0.002023 & 5.54109 & 2.30E-05 & 2.721801 & 0.194735 & 580.71457 & 20.75554 & 1,2,...,13 & &2019-05-07 & 2020-08-03 \\ 
...&1328.046433 & 0.001592 & 10.691632 & 0.000083 & 3.17746 & 0.415448 & 519.241028 & 26.40276 & 1,2,...,13 & &2019-05-17 & 2020-08-03 \\ 
\enddata
\tablecomments{
The TOI catalog table lists the parameters of each TOI and its host star. This table (Part II) gathers together the planetary parameters selected from the TOI catalog with the stellar parameters on a previous page.
}
\end{deluxetable*}
\end{rotatetable*}


\subsection{Known Issues with TOI parameters}

TOI parameters reflect the best possible analysis available at the time of vetting. 
The ``Comments" field in the TOI catalog often records remarks from the vetting team on the reported parameters. 

Common remarks include:
\begin{itemize}
    \item Ambiguous period (twice/half): the flagged transits in the light curve do not clearly determine the correct period. The period may be a multiple or a fraction of the reported period. This often happens when transits are missed because of data gaps. 
    \item Revised epoch: the first transit flagged in the DV Reports is not the first true transit in the light curve and has been revised in the TOI table.
    \item Target period and epoch match with a nearby TCE
    \item Transit is suspected to be due to an eclipse or transit on a nearby star (often with an ambiguous centroid offset)  
    \item Two stars in the same pixel
    \item Low SNR or a marginal candidate
    \item Possible EB or stellar variability
    \item Possible single transit 
    \item Possible grazing transit
    \item V-shaped transit
\end{itemize}

\subsection{Vetted TCE List}

The TOI vetting team also maintains a larger TCE List of SPOC and QLP TCEs considered for vetting, and includes the TOIs. The TCE list has the columns listed below.

\begin{itemize}
    \item {\bf TIC ID}
    \item {\bf Candidate ID}. Identifier for the signal in the light curve (1, 2, etc.)
    \item {\bf Full TOI ID}. Optional, if available.
    \item {\bf Epoch} Barycentric-corrected epoch of first transit in TESS Julian Days. $\text{BTJD}=\text{BJD}-2457000$ \footnote{\citet{SDPDD2018}}. Includes uncertainty.
    \item {\bf Orbital Period} In days. Can be blank, as is the case for single transits. Includes uncertainty.
    \item {\bf TCE Disposition} Optional. 
    \item {\bf Comment} Optional. Describes provenance and reasoning behind the disposition assigned.
    \item {\bf Sector Tag} Identifies in which sector and vetting collection a candidate appeared. 
     \item {\bf Updated} Date of most recent update of candidate parameters.
\end{itemize}

The TCE list is available online alongside the TOI list\footnote{\url{https://tess.mit.edu/toi-releases/}}. Separately, MAST archives all the TCEs the SPOC pipeline finds for each sector and multi-sector run\footnote{\url{https://archive.stsci.edu/tess/bulk_downloads/bulk_downloads_tce.html}}. The SPOC TCE list is part of the public data release for each sector. 
\section{Discussion}
\label{sec-discussion}

\begin{figure*}[ht]
    \includegraphics[width=6.5in]{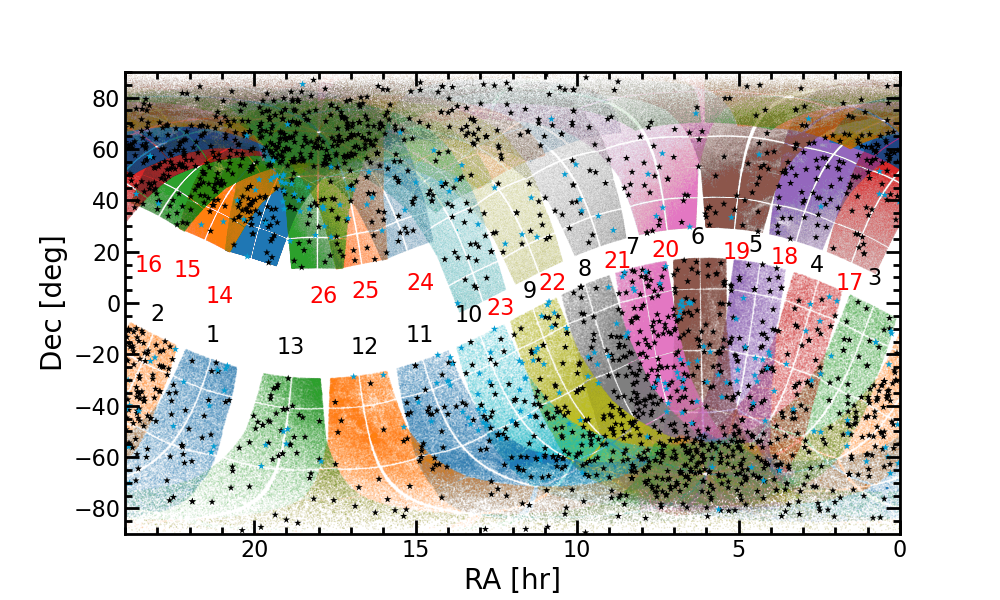}
	\caption{Sky positions of the \tess\ Prime Mission TOIs. RA is on the horizontal axis and declination is on the vertical axis. The black points are TOIs. The blue points are previously known planets (NASA Exoplanet Archive, accessed 14 August 2020). The points within the multi-colored swaths are stars ($T_{\rm mag} < 13.5$) observed in the \tess\ Prime Mission. Fewer TOIs appear in Sectors 11 and 12 because the field of view was crowded and this limited the number of viable TCEs for vetting.}
    \label{fig:RADec}
\end{figure*}

We now turn to a qualitative description of the planet candidates in the TOI list
The TOI list is not corrected for completeness so the conclusions we can draw about the underlying planet populations are limited. Nevertheless, we do present some figures of TOI parameter space distributions.  
The positions of the \tois\ TOIs released in the \tess\ Prime Mission are shown in Figure~\ref{fig:RADec}. 

\subsection{TOI Host Stars}

\begin{figure*}[ht]
    \includegraphics[width=6.5in]{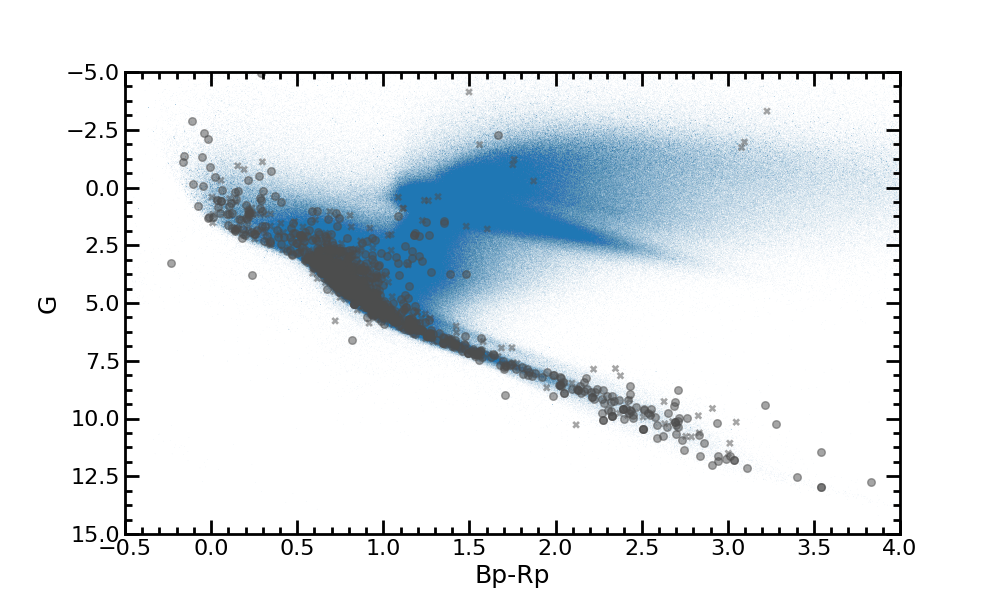}
	\caption{TOI host stars on the Gaia color-magnitude diagram. The horizontal axis is the difference between the Gaia DR2 magnitudes in the $G_{RP}$ and $G_{BP}$ bands and the vertical axis is the absolute G magnitude. The blue points are the bright stars (brighter than \tmag\ = 13.5) \tess\ observed in the Prime Mission matched in Gaia DR2. The same sample of stars is plotted as points in multi-colored swaths in Figure~\ref{fig:RADec}. The grey points are TOIs. The grey X's are TOIs which have been found to be false positives from follow-up observing and revised vetting. The majority of the TOI host stars fall on the main sequence. All the TOIs identified for giant stars brighter than the red clump have proven to be false positives, as expected.}
    \label{fig:toihr}
\end{figure*}

The TOIs are found in a diverse stellar population, from hot A stars to cool M dwarf stars. The majority of the TOI host stars are on the main sequence, while some are sub-giants (Figure~\ref{fig:toihr}). 

\begin{figure}[ht]
    \includegraphics[width=3.4in]{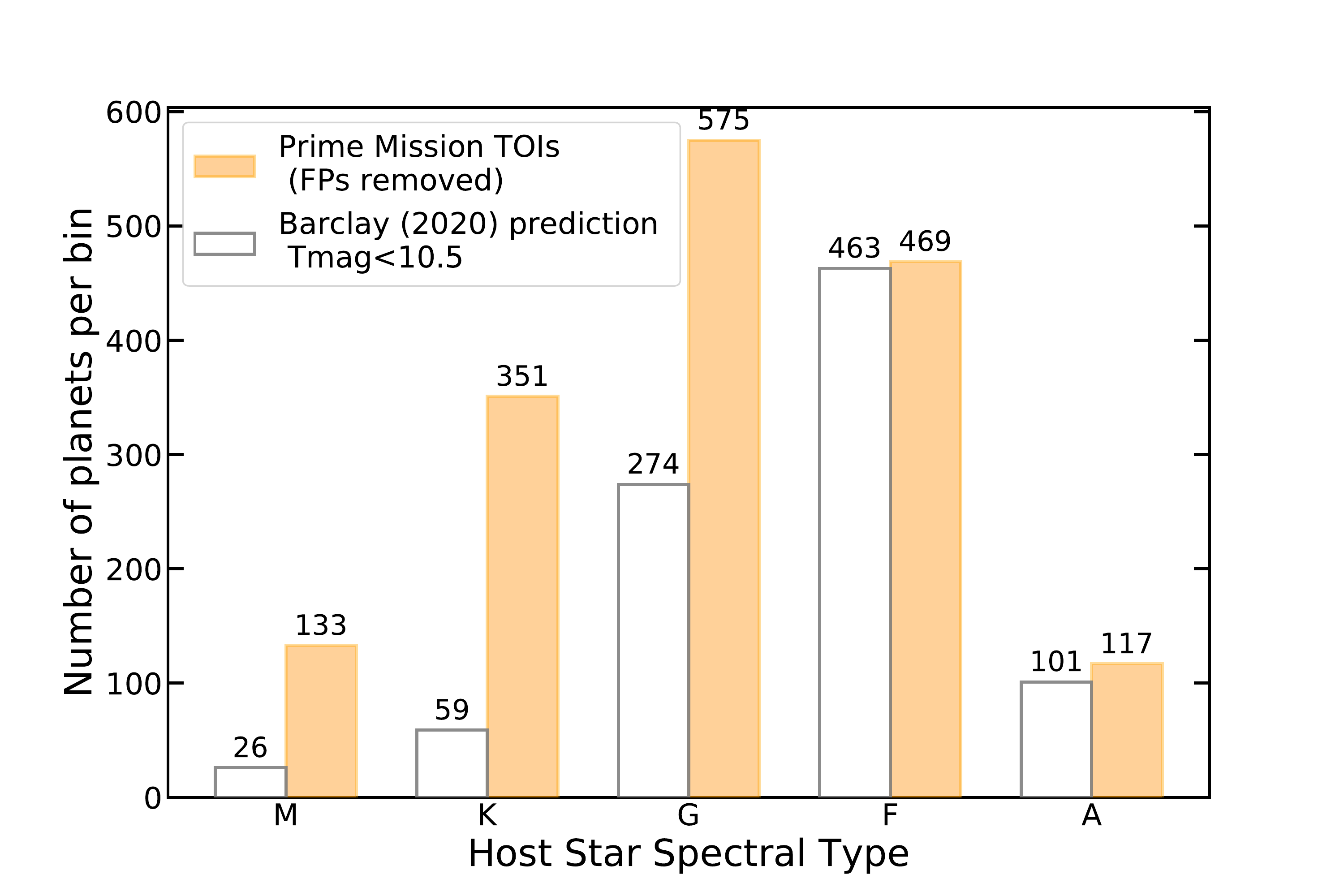}
    \caption{Spectral type distribution for TOI host stars. The number of TOIs for each spectral type (M,K,G,F,A) is on the vertical axis. The grey bars are the \citet{barclay2020} predicted yield for each spectral type for stars brighter than $\tmag = 10.5$. The orange bars are the numbers of TOIs actually identified in the \tess\ Prime Mission for each spectral type. False positive TOIs have been removed. The TOIs roughly track to the abundance of each spectral type, but the data are not corrected for completeness.}
    \label{fig:spectbar}
\end{figure}

\begin{figure*}[ht]
    \includegraphics[width=6in]{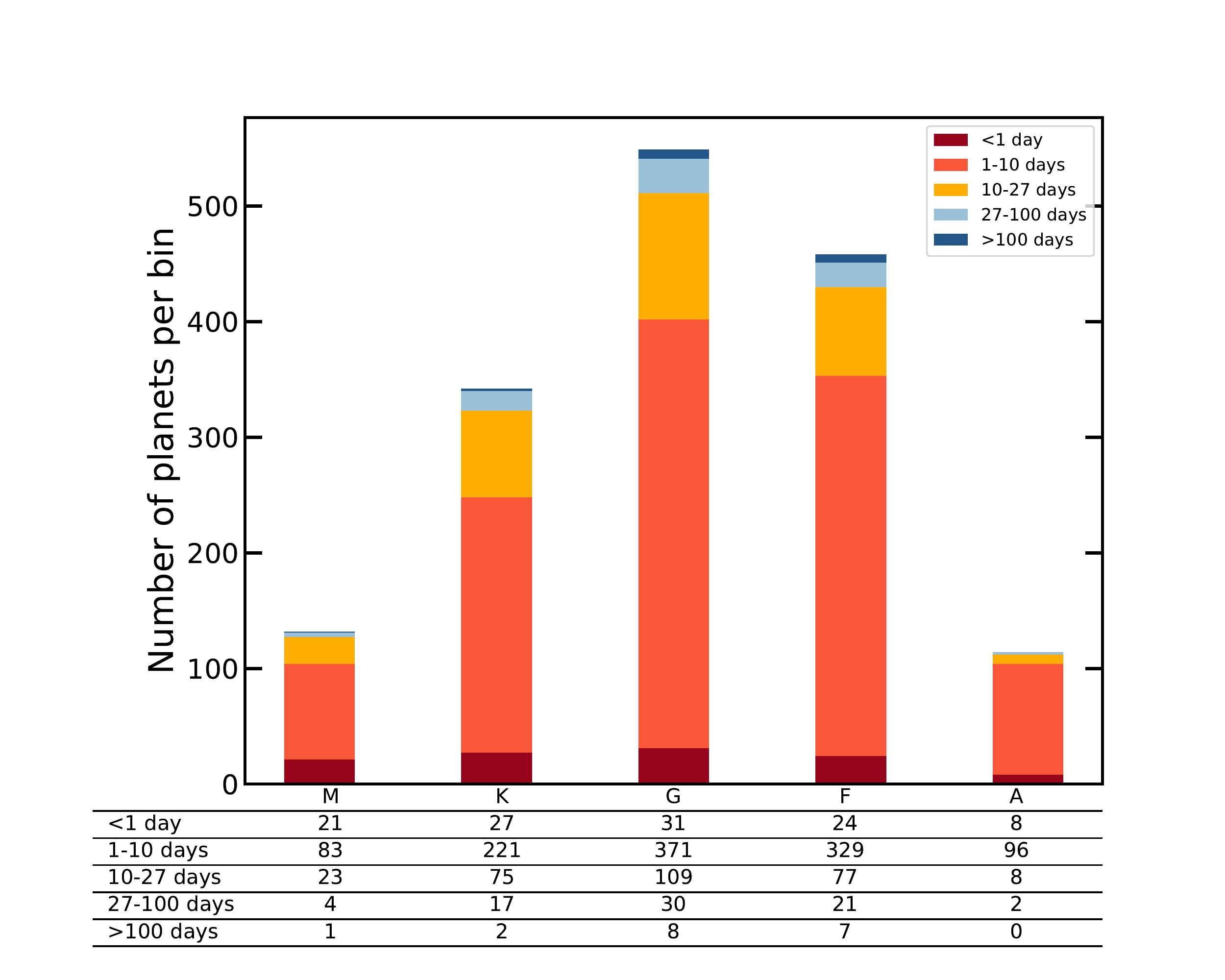}
    \caption{Period distribution for TOIs, subdivided by host star spectral type (M,K,G,F,A). False positive TOIs and TOIs with $P=0$ are not included. For each spectral type, \tess\ has detected many TOIs in the 1-10 day period regime. These data are not corrected for completeness.}
    \label{fig:spectper}
\end{figure*}

\begin{figure*}[ht]
    \includegraphics[width=6in]{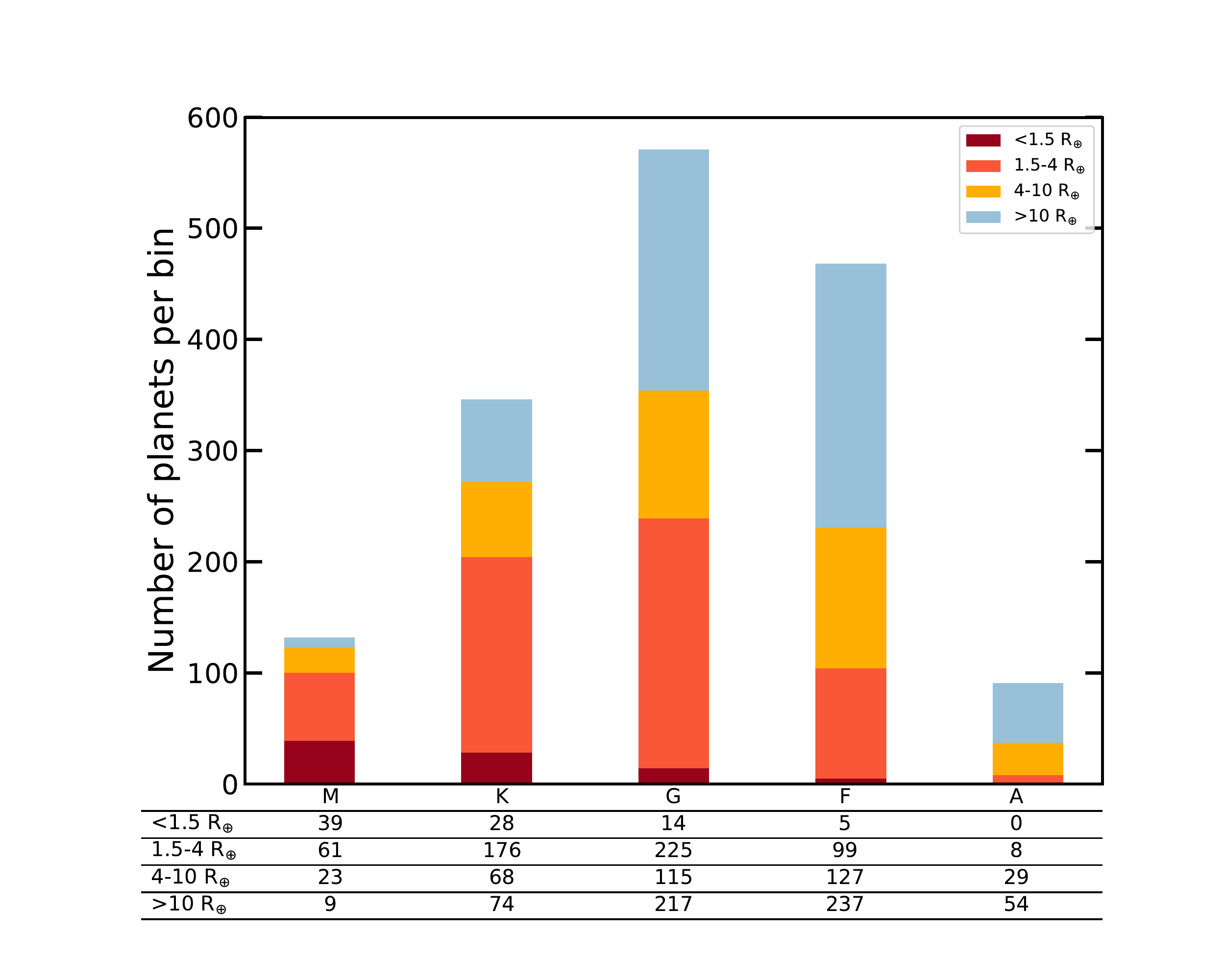}
    \caption{Radius distribution for TOIs, subdivided by host star spectral type (M,K,G,F,A). False positive TOIs and TOIs with no $R_{p}$ are not included. The majority of TOIs on M and K-type stars are sub-Neptunes, demonstrating \tess's ability to detect small planets around small stars. These data are not corrected for completeness.}
    \label{fig:spectrad}
\end{figure*}

TOIs appear around primarily FGK stars, as shown in the spectral type distribution in Figure~\ref{fig:spectbar}. Short-period TOIs predominate across all spectral types, as shown in Figure~\ref{fig:spectper}. Figure~\ref{fig:spectrad} shows that for M and K-type stars, small TOIs (Neptune-sized and smaller) predominate, whereas for F, and A-type stars, TOIs with large radii are more common. This distribution is expected, as the transit depth \tess\ measures for a planet of a certain radius is shallower for larger stars than for smaller stars; small planets are more difficult to detect around larger stars.

\subsection{TOI Planet Parameters}
\begin{figure*}[ht]
        \includegraphics[width=6.5in]{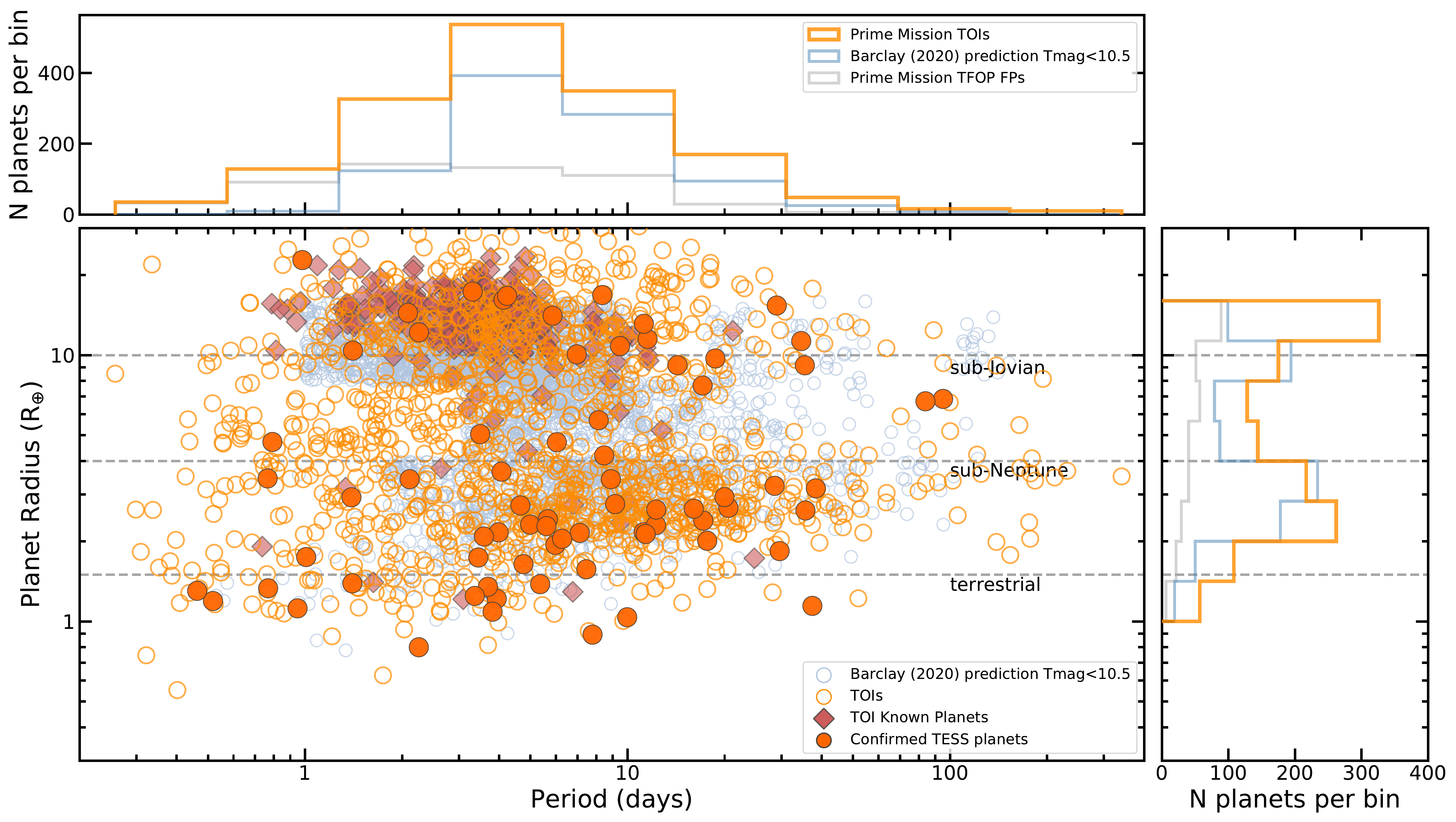}
	\caption{TOIs in period-radius space. The period in days is on the horizontal axis and the planet radius in Earth radii is on the vertical axis. The solid dark orange circles are \tess\ planets confirmed either with a mass measurement or another validation method. The red diamonds are known planets from other surveys re-detected by \tess. The open orange circles are TOI planet candidates excluding confirmed \tess\ planets, known planets, and false positives. The light blue open circles are the predicted \tess\ planet yield from \citet{barclay2020} for stars brighter than $\tmag = 10.5$. The banding in the simulated points, for example at $4 \rearth$, is an artifact of the period-radius bins in which occurrence rates were calculated.
	The top histogram shows the distribution of TOIs in period space. The histogram at right shows the distribution of TOIs in radius space. The TOIs, including known planets and confirmed \tess\ planets, are the orange histogram. The predicted planet yield from \citet{barclay2020} is in light blue, and the false positives as of \accessed\ are in grey.
	}
    \label{fig:periodradius}
\end{figure*}

 In Figure~\ref{fig:periodradius}, we plot the TOIs in period-radius space. We include the simulated planet distribution from \citet{barclay2020} to compare to the TOI catalog. We only include predicted planets with $\tmag < 10.5$ host stars, as the vetting team only searches for TOIs for QLP targets above this magnitude threshold.
 The period and radius distributions for the TOIs, the \citet{barclay2020} prediction, and the false positives from follow-up as of \accessed\ are shown at the top and right of Figure~\ref{fig:periodradius}.
 There are 820 TOIs with $R_p < 6 \rearth$ out of 1625 with calculated planet radius values. The majority of the TOIs have periods less than 30 days, which is expected considering the single-sector baseline is 27 days. The median period is 4.5 days and the median planet radius is 5 \rearth.
 The TOIs appear to be in in good qualitative agreement with the predicted yield estimations by \citet{barclay2020}. 
 The majority of the false positives (plotted in grey in Figure~\ref{fig:periodradius}) appear at $P < 6 $ days and $R_{p} > 5$ \rearth. This is expected due to the contribution of nearby or on-target eclipsing binaries. 
 
\begin{figure*}[ht]
        \includegraphics[width=6.5in]{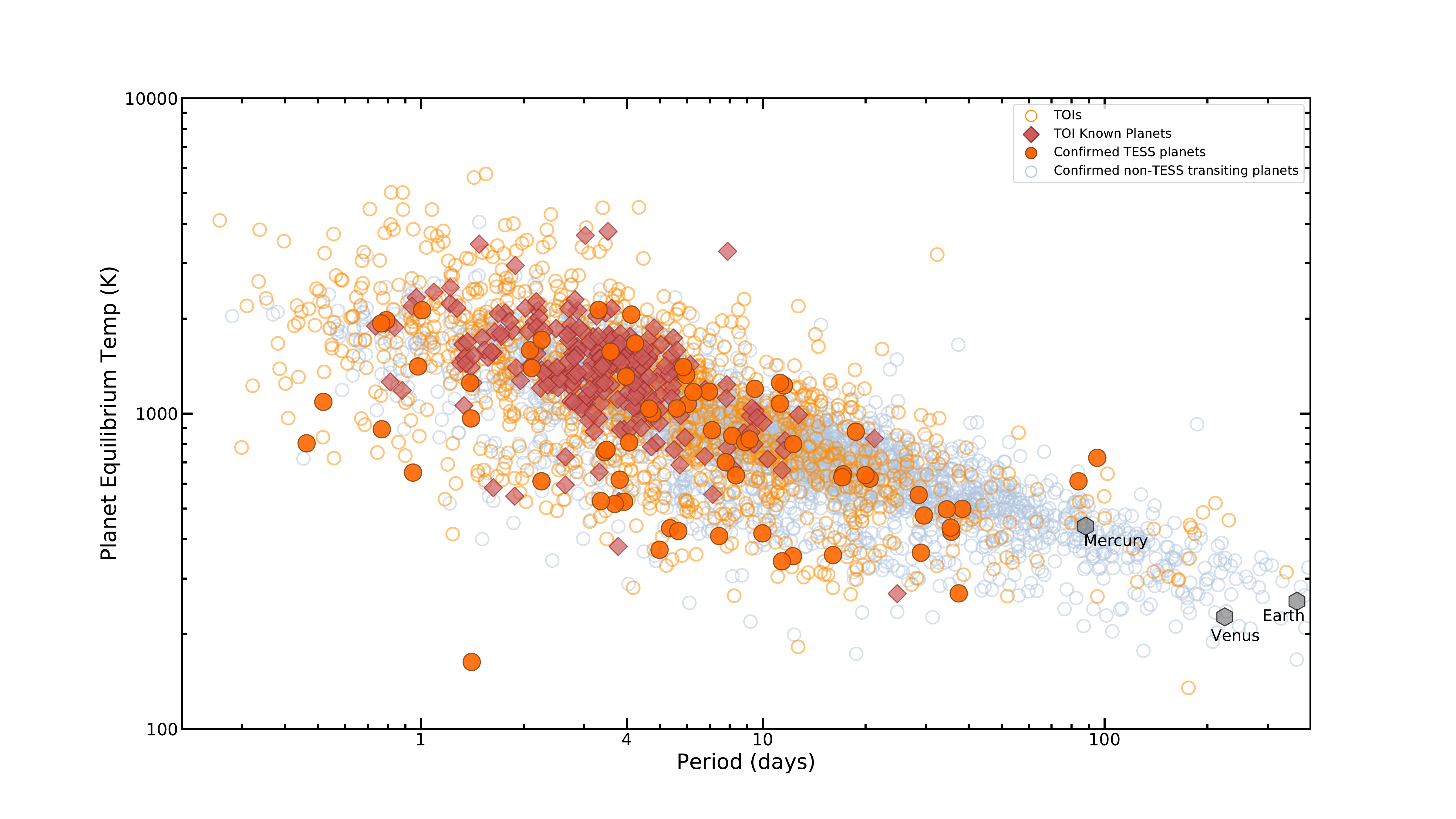}
    \caption{Orbital period vs. planet equilibrium temperature. The planet orbital period in days is on the horizontal axis and the planet $T_{eq}$ is on the vertical axis. The open orange circles are planet candidate TOIs which have not been confirmed. The solid dark orange circles are planets confirmed either with a mass measurement or are validated by other means. The red diamonds are TOIs which were known planets prior to the TESS mission. The light blue open circles are known planets from outside the \tess\ mission (NASA Exoplanet Archive, accessed \accessed). Mercury, Venus, and Earth are visible in the bottom right. The TOI catalog has greatly increased the number of temperate, short-period planet candidates available for follow-up.} 
    \label{fig:periodvsteq}
\end{figure*}

Looking at the TOIs in period-$T_{eq}$ space in Figure~\ref{fig:periodvsteq}, we see that the TOI Catalog and \tess\ confirmed planets are contributing to the population of temperate planets orbiting their host stars at short periods ($P < 10$~days). 

\subsection{Follow-Up Observations of TOIs}
\label{subsec-tfop}

\begin{figure*}[ht]
    \includegraphics[width=6.5in]{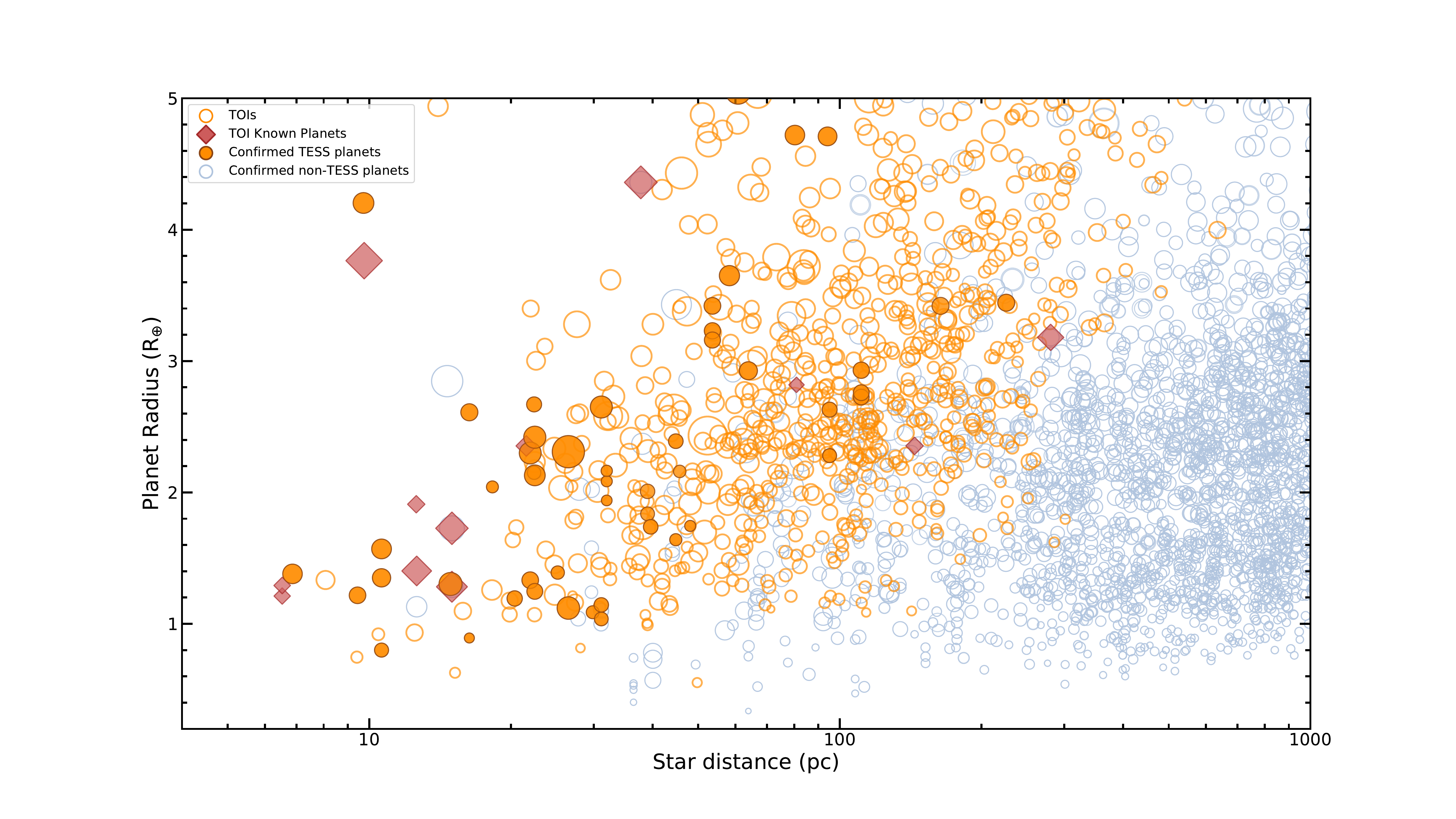}
	\caption{Planet radius (in $\rearth$) vs.~host star distance (in pc). Solid dark orange circles are \tess\ planets confirmed either with a mass measurement or other validation method; orange open circles are TOIs (with known false positives removed); the solid red diamonds are known planets from other surveys re-detected by \tess; and blue open circles are known exoplanets from outside the TESS mission (NASA Exoplanet Archive, accessed \accessed). Point size is scaled by transit depth.}
    \label{fig:RvsD}
\end{figure*}

\TESS\ aims to provide small planets around bright stars that can be characterized in-depth via follow-up observations within three years of the survey onset. 
\TESS\ is delivering on its promise, by discovering many small planet candidates within 100 pc of the solar system (see Figure \ref{fig:RvsD}). Over the course of the \TESS\ Prime Mission, \tois\ TOIs have been sorted through by follow-up observers, 654 of which are planet candidates with radii smaller than $4 R_{\Earth}$, so there will be no shortage of viable targets for mass measurements.   

\begin{figure*}[ht]
    \includegraphics[width=6.5in]{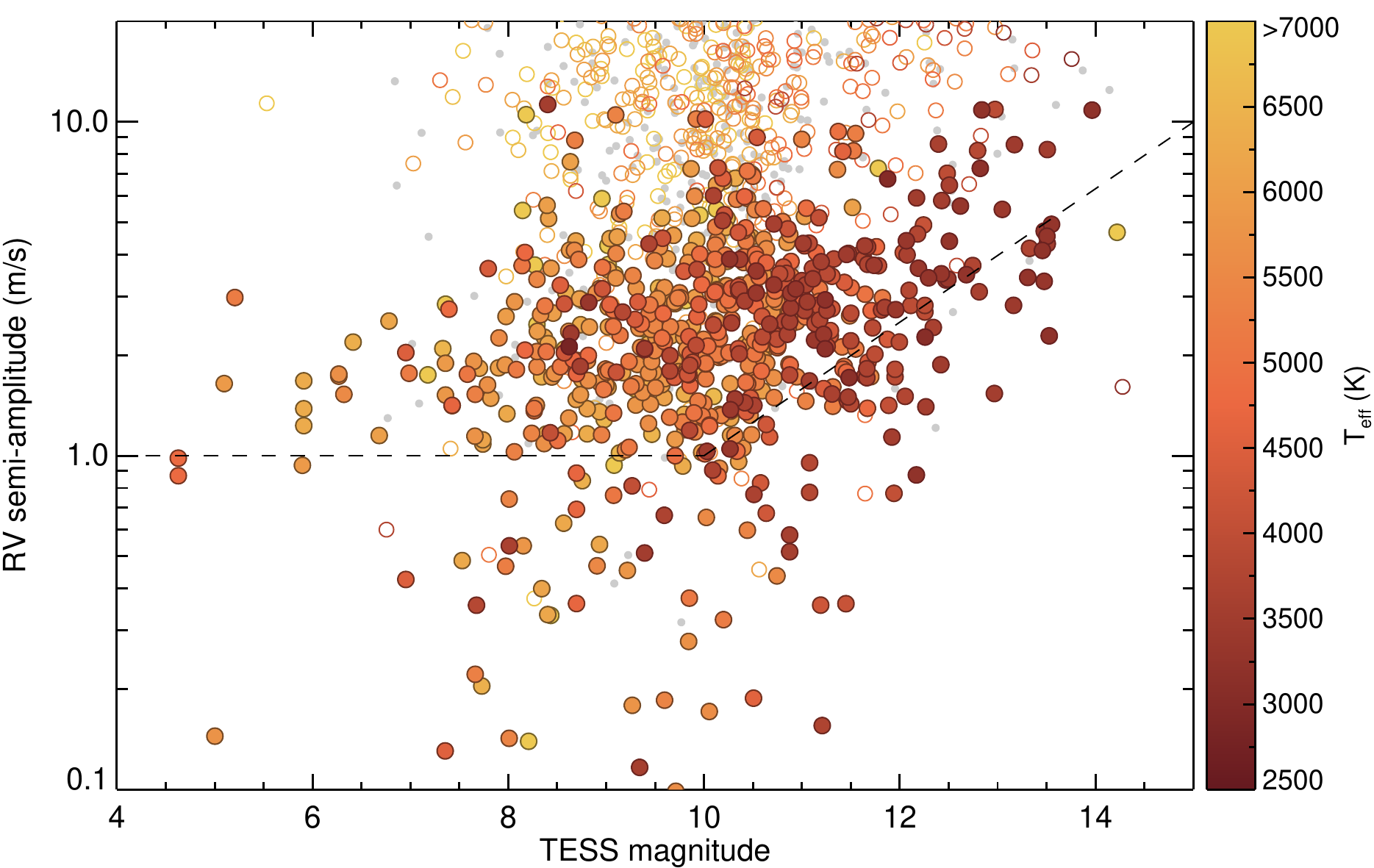}
	\caption{Estimated RV semi-amplitudes for TOIs. RV semi-amplitude (in meters per second) is on the vertical axis and TESS magnitude is on the horizontal axis. False positives (of all sizes) from follow-up and revised vetting are small gray points in the background. Planets (both TESS confirmed and known planets) and TESS planet candidates which are smaller than $4 \rearth$ are filled circles; those larger than $4 \rearth$ are open circles. The color bar is host star \teff. The semi-amplitude calculation uses estimated planet masses derived from radii based on the probabilistic Chen and Kipping mass-radius relation \citep{chen:2017} and uses stellar masses and radii from TICv8 when available. For radii not in TICv8, we use Gaia DR2 or assume 1 $R_{\odot}$. 
	If the stellar mass is not in TICv8, it is approximated to equal the radius of the star, with the exception of evolved stars, for which we assign a maximum stellar mass of 1.5 $M_{\odot}$. The dashed line indicates an approximate threshold for measurable RV signals, using 1 m/s as the lower limit for detectability, increasing with photon noise for fainter host stars.
	Most TESS candidates fall above this threshold for mass measurement, but we note that in practice, stellar activity correction will be needed for many of the smallest signals.
	}
    \label{fig:toi_rv}
\end{figure*}

The majority of the TOIs in the catalog are accessible to RV follow-up, as shown in Figure~\ref{fig:toi_rv}. To date, 25 planets\footnote{\label{fn:exofop}NASA Exoplanet Archive, accessed \accessed} discovered by \TESS\ smaller than 4 \rearth\ have published mass measurements in peer-reviewed publications
and dozens more such measurements are underway. 
An additional 21 larger planets\footnote{\ref{fn:exofop}} have measured masses, again with many more mass measurements underway. 
With numerous new planet candidates and confirmed planets added to the TOI catalog in the Prime Mission, TESS has made progress towards its primary goal, to deliver fifty small ($R_{p} < 4 \rearth$) planets with measured masses to the community. 

\begin{figure*}[ht]
    \includegraphics[width=6.5in]{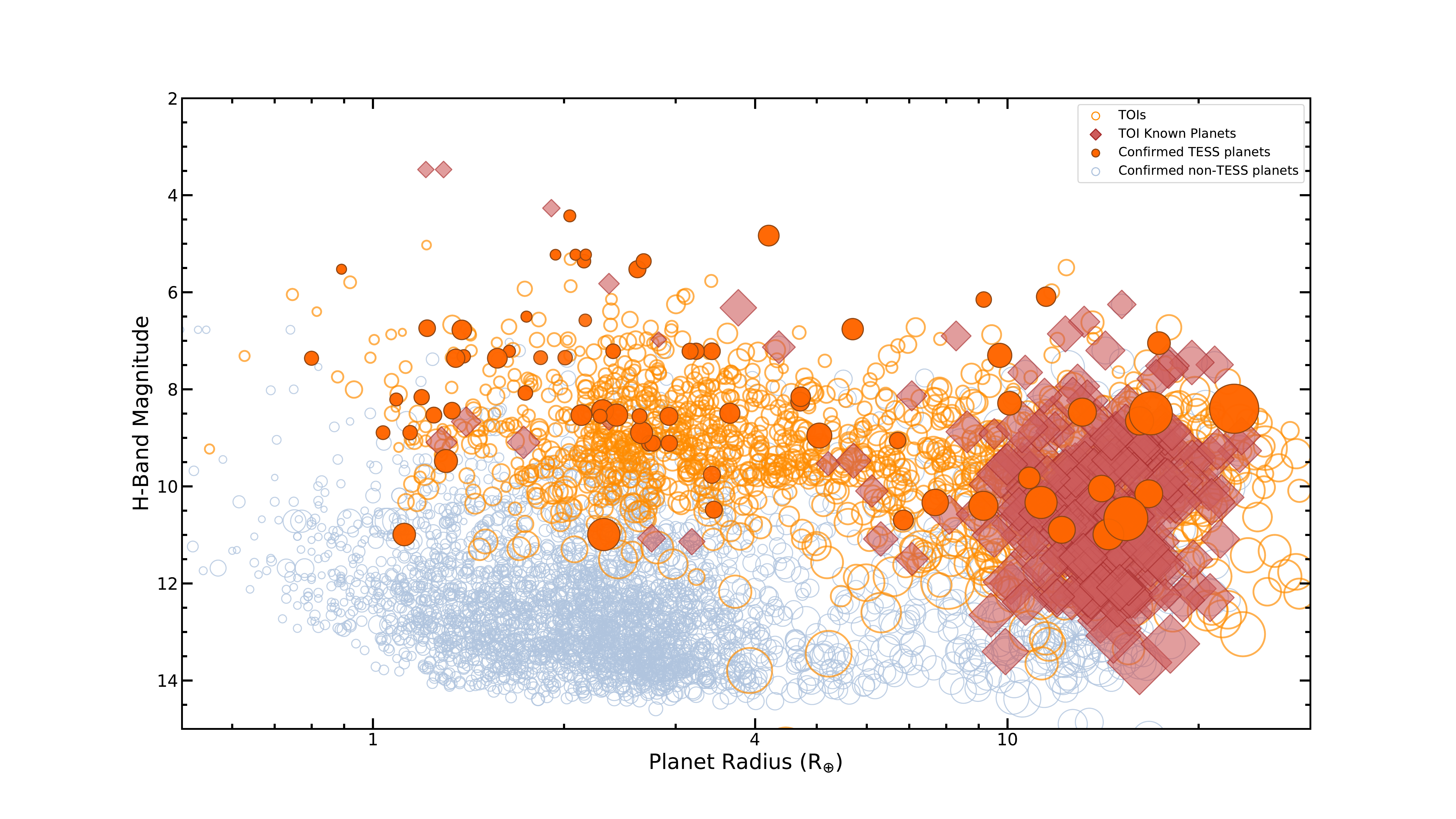}
	\caption{Host star magnitude (in H$_{\rm mag}$) vs.~planet radius (in $R_{\Earth}$). The solid dark orange circles are \tess\ planets confirmed either with a mass measurement or validated by other means; the open orange circles are TOIs (non-validated, but with known false positives removed); the solid red diamonds are known planets from other surveys re-detected by \tess; and the pale blue open circles are known exoplanets (dominated by validated planets, i.e. without mass measurements) from outside the \tess\ mission (NASA Exoplanet Archive, accessed \accessed). The point size is scaled by transit depth. \tess\ is discovering planet candidates with bright host stars amenable to follow-up observations. We plot the host star's  $H_{\rm mag}$ because currently, atmospheric follow-up with the \hubble\ Wide Field Camera 3 instrument is often carried out in the H band.}
    \label{fig:RvsHmag}
\end{figure*}

The TOI catalog also identifies planets which will be candidates for in-depth atmospheric characterization with transmission and emission spectroscopy. 
Many of the small planets TESS discovered have host stars with apparent H-band magnitudes brighter than 9 (Figure~\ref{fig:RvsHmag}) and for fainter host stars, have large transit depths.

\begin{figure*}[ht]
    \includegraphics[width=6.5in]{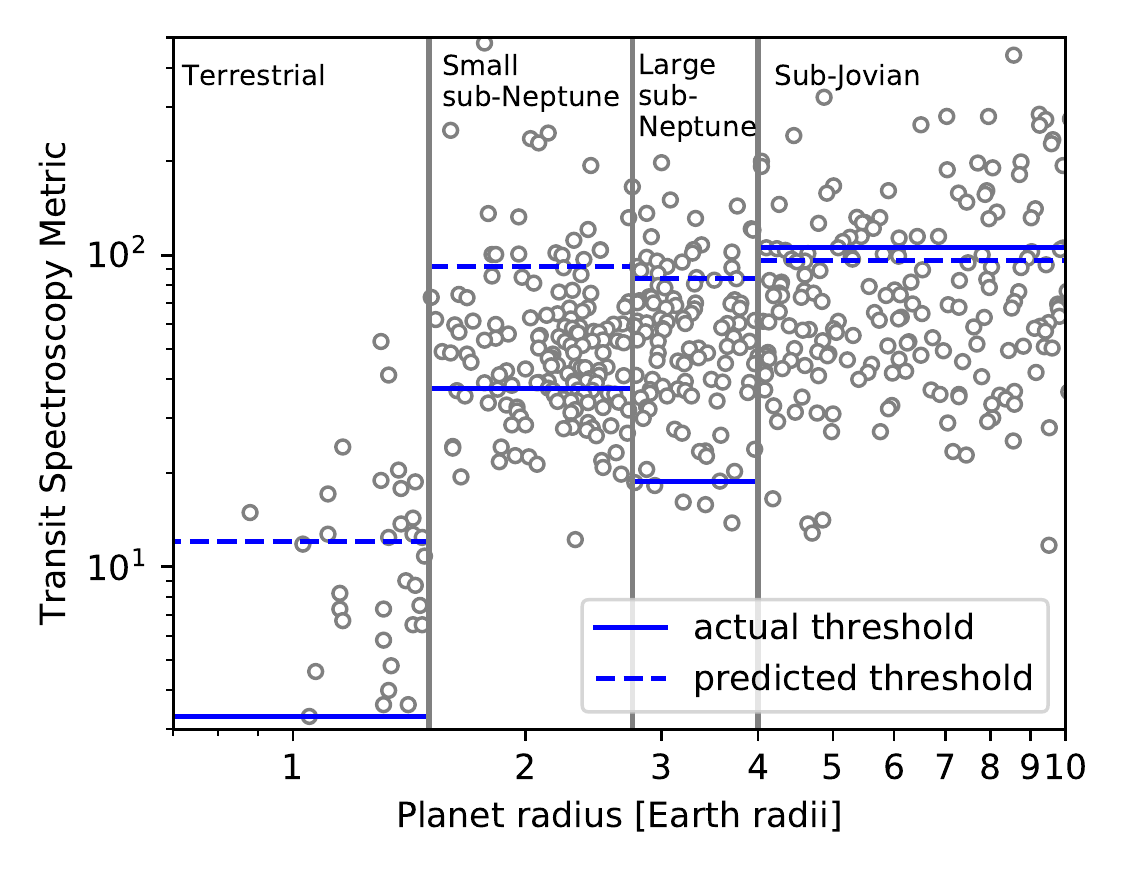}
	\caption{The Transmission Spectroscopy Metric (TSM) \citep{kempton2018} as a function of planet radius for \tess\ candidate planets from the Prime Mission (including known planets, but excluding false positives). The solid blue lines indicate the threshold TSM required to obtain a sample of 37 terrestrial planets, 100 small sub-Neptunes, 100 large sub-Neptunes, and 50 sub-Jovians over the entire \tess\ mission.  The dashed blue lines indicate the predicted threshold TSM from \cite{kempton2018}. These results suggest that the TSM values  for \tess\ sub-Neptunes ($R_p<4 R_\oplus$) are lower than predicted by \citep{kempton2018}, and that more  observing time than expected may be required to build a statistical sample of planets in each category. The difference between the actual and predicted values, however, may be in part due to an overprediction of short-period planets in the yield simulations. The actual TSM values may shift as the \tess\ data continue to be vetted and more TOIs are found.}
    \label{fig:toi_tsm}
\end{figure*}

Figure \ref{fig:toi_tsm} shows the transmission spectroscopy metric (TSM) \citep{kempton2018} values for TOIs in a range of radius bins. The TSM is the signal-to-noise expected for spectral features in a transmission spectrum. \cite{kempton2018} estimated TSM threshold values that would provide a statistical sample of the best TESS targets across the full radius range for the \jwst\ observations, based on the predicted \tess\ yield from \cite{sullivan2015}. For a quantitative calculation of the observing time needed to reach a certain TSM with \jwst, see \cite{kempton2018, louie2018}. Figure \ref{fig:toi_tsm} shows updated TSM threshold values  from the Prime Mission TOI catalog. Although the updated TSM thresholds are lower than predicted for sub-Neptunes (with radii below $4 \rearth$), this may be in part due to an overprediction of the number of short-period planets in yield estimations. With the current TOI catalog, more observing time than expected may be required to characterize a statistical sample of small planets. However, additional planets may be found with further vetting. The highest TSM planets have already provided a rich source of targets suitable for atmospheric characterization with \hubble: pi Men c \citep{2020hst..prop16065K}; LTT 1445A b \citep{2019hst..prop16039S}; TOI 1726.01 and 1726.02 \citep{2020hst..prop16319Z}; AU Mic b \citep{2019hst..prop15836N,2020hst..prop16164C}; TOI-674 b \citep{2017hst..prop15333C}; DS Tuc A b \citep{2020hst..prop16085D}; HD 219666 b \citep{2019hst..prop15698B,2019hst..prop15969F}; TOI-270 c and d \citep{2019hst..prop15814M}; TOI-1231b \citep{2020hst..prop16181K}; and LP 98-59 b, c, and d \citep{2019hst..prop15856B}.


\subsection{Previously-Known Exoplanets in the TOI Catalog} 
\label{subsec-kps}

By systematically observing the sky, \tess\ is monitoring many of the stars already known to harbor exoplanets (blue stars in Figure~\ref{fig:RADec}). For known transiting exoplanets, \tess\ provides improved transit ephemerides. The improvements reduce the transit timing uncertainty in the \jwst\ era from tens of minutes down to a few minutes. The TOI catalog includes 256 Known Planet TOIs observed in the Prime Mission.

The known exoplanet host stars were included as a special catalog in the originally released version of the TIC \citep{stassun2018, stassun2019}. Several selected programs from the \tess\ Guest Investigator Program for Cycle 1 and Cycle 2 included known exoplanet host stars, ensuring that they would be monitored with two-minute cadence by \tess. 

Using data from the NASA Exoplanet Archive\footnote{Accessed \accessed} \citep{akeson2013} and cross-matching with the \tess\ camera locations, we estimate that $\sim920$ known hosts of transiting exoplanets brighter than \tmag\ = 13.5 were observed by \tess\ during the Prime Mission.
 We detected 251 of the 1,543 total known planets around $\tmag < 13.5$ host stars originally detected via the transit method.  This smaller number is expected, as many of these known planets were discovered by \Kepler\ and the SNR for transits of many of these planets is too low for \tess. We released TOIs for 5  of the 864 known planets detected via other methods and orbiting host stars with $\tmag < 13.5$. We expect the number of TOIs initially detected by radial velocity measurements to be low because of low transit probabilities for radial velocity targets \citep{dalba2019,kane2021}.

\subsection{Multi-Planet Systems and Multi-Planet False Positive Probability}

\begin{figure}[ht]
    \includegraphics[width=3.5in]{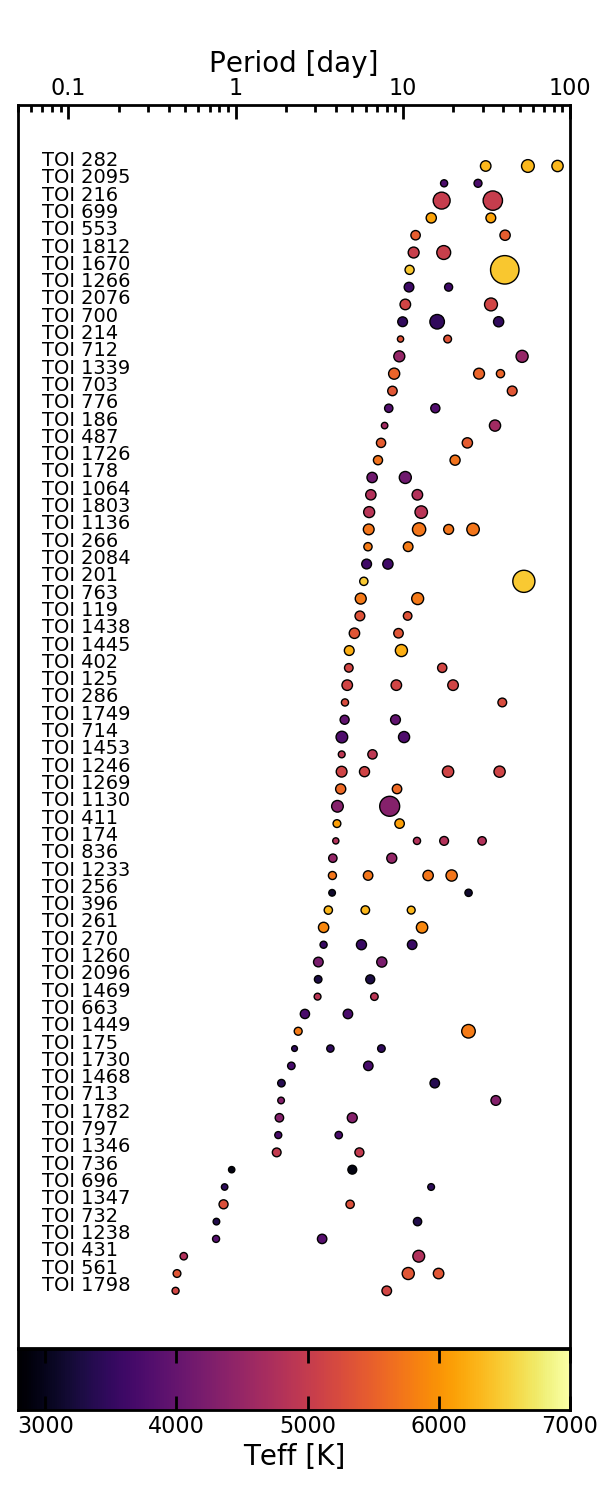}
	\caption{Multi-planet TOI systems. The horizontal axis is period in days, and the TOIs are ordered from longest to shortest minimum period. The color of each planet marker maps to the effective temperature of each system's host star and the planet markers are scaled by planet radius. The multi-planet systems \tess\ has discovered have host stars across a wide range of spectral types.}
    \label{fig:multisystems}
\end{figure}

The TOI planet candidates includes several dozen systems containing up to four planet candidates each. This is despite \TESS's shorter observing baseline compared to \Kepler\ and K2. The architectures of the TOI multi-planet systems are summarized in Figure \ref{fig:multisystems}.  

Early on in the \Kepler\ mission, \citet{latham2011,lissauer2011} noticed that planet candidates in systems with multiple transiting planet candidates (i.e. multi-transiting systems) were considerably more likely to be genuine transiting exoplanets than those in single planet candidate systems (i.e. singly-transiting systems). This observation was formalized into a statistical framework \citep{lissauer2012} and used in the validation of hundreds of planetary candidates in multi-transiting systems \citep{lissauer2014, rowe2014}. Here, we investigate whether multi-transiting systems discovered by \TESS\ are similarly more likely to be genuine transiting planets than their singly-transiting counterparts.

We followed a procedure similar to \citet{lissauer2012} to determine how much more likely multi-transiting candidates are to be true positives than singly-transiting candidates. In particular, we calculated a ``multiplicity boost'' - a multiplicative factor that can be applied to the false positive probability of any given transiting planet candidate to reflect the {\em a priori} higher likelihood of a candidate in a multi-transiting system to be a real planet. 

We consider only TOIs found by the SPOC pipeline because the search target list for these TOIs is well-defined. Over the Prime Mission, \TESS\ observed 232,705 unique stars for at least one sector at two-minute cadence. From the stars observed at two-minute cadence, the TOI process identified 1023 total planet candidates in 930 systems. There were 855 single-planet-candidate systems, 61 systems with two planet candidates, 10 with three candidates, and 4 with four candidates. 

The multiplicity boost is given by the ratio of probability any star is a TOI and the probability that any TOI has more than one planet candidate. In this case, the multiplicity boost is roughly (75/930) / (930/232705) $\approx 20$. This is lower than the multiplicity boost estimated from both \Kepler\ and K2 \citep[about 30,][]{lissauer2012, sinukoff, vanderburg16d}, likely because of \TESS's higher false positive rate (due to larger pixels), and the increased difficulty in detecting and disentangling multi-transiting systems given \TESS's shorter observing baselines.  



The multiplicity boost is stronger when we restrict our analysis to only planets smaller than 6 \rearth, which are intrinsically more common and therefore have lower false alarm rates. From the two-minute postage stamps, we identified 630 planet candidates smaller than 6 \rearth in 545 systems. The majority of \tess\ multi-planet systems (69/76) have more than one small planet candidate: 56 systems with two planet candidates, 9 with three, and 4 with four planet candidates. Calculating the multiplicity boost for this set of smaller planet candidates gives (69/545)/(545/232705) $\approx$ 54. The presence of multiple small planet candidates is a strong indicator that the system is genuine. 

We advise caution when applying the multiplicity boost to candidate multi-transiting systems in crowded regions like the galactic plane because the likelihood of many false positives in the same pixel is higher. Our multiplicity boost was calculated as an average over the whole sky but is likely lower in crowded regions. We also note that these estimates for the multiplicity boost do not apply to planet candidates detected only by the QLP. Because the QLP searches all stars observed in the full frame images, regardless of their evolutionary state and suitability for detecting transiting planets (such as red giants), QLP planet candidates have a higher false positive rate. Any multiplicity boost present for multi-planet systems detected only by the QLP will likely be much weaker than the boost for SPOC pipeline-detected multi-planet systems. So far, the QLP has only detected one system with multiple candidates (TOI 1130, \citealt{huang1130}) for which the stars were not pre-selected for two-minute cadence observations and searched by the SPOC pipeline, so we cannot estimate a multiplicity boost for candidates only detected by the QLP. 

\subsection{Single Transits}
\label{subsec:single-transits}

Planet candidates which transit only once during \TESS\ observations are of interest because they can be followed up to yield rare, long-period transiting planets. The QLP and SPOC pipelines search for at least two transits to make a TCE and therefore do not purposefully generate a list of single-transit planet candidates. However, occasionally the pipeline finds a single transit and folds it on a transit-like noise feature elsewhere in the light curve. When this happens, we include these single transit events in the TOI Catalog with $P=0$.  
It is possible for a TOI which is a single-transit planet candidate with a single sector of data to be updated with a measured period if additional transits are found in later sectors.   

There are separate, dedicated efforts to purposefully identify and follow-up single transits from the QLP and SPOC and also from these independent searches. In the latter case of single-transit planet candidates from independent searches, these candidates fall into the category of community TOIs. 

The follow-up efforts for single-transit planet candidates fall into two categories. The first is photometric follow-up of a list of target stars with single transits \citep[for example][]{cooke2018}. 
The second is a dedicated radial-velocity campaign for a list of target stars, using an initial estimate of the period \citep{Seager2003} by using transit and stellar properties and assuming a circular orbit (e.g., Villanueva et al., in prep.).

\subsection{Some Notable TESS Firsts}

Here we present a few highlights of TESS exoplanet discoveries, selected because they are \tess\ ``firsts'' or appear in an otherwise scarce parameter space in stellar parameters or planet parameters.

\paragraph{$\pi$ Mensae c} The mission's first planet discovery \citep{huang2018}, $\pi$ Men c (TOI 144), delivered on \tess's promise to find small planets around bright stars. $\pi$ Men c orbits a very bright host star ($\vmag=5.7$). The planet is a sub-Neptune with an average density too low ($M = 4.82_{-0.86}^{+0.84}~M_{\oplus}$, $R = 2.042 \pm 0.050  ~R_{\oplus}$,  $\rho = {2.97}_{-0.55}^{+0.57}$ $\rho_{\oplus}$) for the planet to be predominately rocky. Instead the planet likely has a thick envelope. The quick $\pi$ Mensae c mass determination was possible due to a rich radial-velocity data set already archived \citep{jones2002}. The system was observed in multiple sectors in \tess\ Year 1. Additional observations and analysis have provided more precise mass measurement for planets b and c, their large orbit misalignment, and possible theories for the formation of the system \citep{kane2020,xuan2020,kunovac2020,damasso2020,derosa2020}. Although $\pi$ Men is a bright star, it is a viable candidate for \jwst\ observation.

\paragraph{TOI 125} TOI 125 is a three-planet system \citep{quinn2019} and was \TESS's first multi-planet system released. The system contains three sub-Neptune planets, and two possible additional small planets detected in further analysis with low ($5.2$-$\sigma$ and $5.1$-$\sigma$) signal-to-noise. The system orbits a K0 dwarf (V=10.9).

\paragraph{TOI 700 d} TOI 700 d is the first \tess\ discovery of an Earth-sized planet in the habitable zone of its star. The host star is an M2 red dwarf star. The planet is  $1.144_{-0.061}^{+0.062} \rearth$ with a period of $P=37.42$ days and an equilibrium temperature of $T_{eq}\sim 269$~K \citep{toi700rodriguez,toi700gilbert,toi700suissa}. The system has two other Earth-sized planets interior to TOI 700 d $(R_{b}=1.037_{-0.064}^{+0.065} \rearth$ and $R_{c}=2.65_{-0.15}^{+0.16} \rearth)$ and lies in the southern \tess\ continuous viewing zone. The system was one of the last targets \emph{Spitzer} observed.

\paragraph{TOI 1338} TOI 1338 b is the first \tess\ circumbinary planet \citep{toi1338kostov}. The planet orbits two stars with masses $1.1 M_{\odot}$ and $0.3 M_{\odot}$ and a 14.6-day orbit. The planet has a nearly-circular orbit, an orbital period of 95.2 days and a radius of $6.85 \pm 0.19 \rearth$. The discovery relied on visual inspection to identify the planet transits as distinct from the secondary transits of the eclipsing binary.

\paragraph{LHS 3844 b} The first ultra-short-period (USP) planet from \tess, LHS 3844 b \citep{vanderspek19} is a hot super-Earth orbiting its host star every 11 hours. The M dwarf host star is 14.9 parsecs away. This candidate demonstrated \tess's ability to detect small planets around nearby small stars. LHS~3844~b was later found to likely be a bare rock planet with little or no detectable atmosphere \citep{kreidberg19}. 

\paragraph{HD 21749} TOI 186.01, or HD 21749 b, is \tess's first long-period  planet, with $P=35.61$ days \citep{toi186b}.  The host star is a bright (\vmag=8.5) K dwarf at 16 parsecs distance. 
TOI 186.01 was initially a single-transit TOI in Sector 1 until additional transits occurred in following sectors. A second, shorter-period planet, TOI 186.02, or HD 21749 c, (P=7.9 days) appeared in multi-sector analysis of SPOC pipeline postage stamp data from Sectors 1-3. TOI 186.02 is the  first Earth-sized planet ($R_p = 0.892 \rearth$) from \tess.
HD 21749b is sub-Neptune-sized ($R_p = 2.61_{-0.16}^{+0.17} \rearth$).
The two planets are on either side of the radius valley \citep{fulton}. 

\paragraph{TOI 200.01} DS Tuc A b \citep{newton19} is \tess's first young  planet ($\tau \sim 45$~Myr). The host star, DS Tuc A, is part of a bright (\vmag=8.5) visual binary in the young ($\tau \sim$~45 Myr) Tucana-Horologium association. The planet is $5.86 \pm 0.17 \rearth$ with an 8.1 day period, a rare size-period combination. This young planet can be a point of comparison against older planets with future measurements of its mass and atmosphere.

\paragraph{TOI 704} TOI 704 b, or LHS 1815 b, is TESS's first thick disk star with a transiting planet \citep{gan20}. In other words, TOI 704 b's host star has a much higher expected maximal height ($Z_{max}$ = 1.8 kpc) above the Galactic plane as compared to other TESS planet host stars. This TOI will enable studies on whether interior structure and atmospheric properties of a thick disk planet differs from planets in the thin disk (or Galactic halo).
TOI 704 b  transits an M dwarf star with
$P=3.8$ days and radius $1.088 \pm 0.064 \rearth$ with a $3-\sigma$ mass upper limit of 8.7 $\mearth$. 

\paragraph{TOI 813} Members of the public can label transit-like features of \tess\ light curves on the Planet Hunters \tess\ platform\footnote{\url{https://www.planethunters.org}}. TOI 813 b was the first \tess\ planet discovered by Planet Hunters \tess\ \citep{toi813eisner}. The Planet Hunters TESS team identified the candidate and released it as a community TOI. The planet orbits a bright, sub-giant star at a long orbital period of
83.4 days and has a radius of $6.71 \pm 0.38 \rearth$.

\subsection{Some Notable TESS Exoplanets}
In addition to being \tess\ firsts, a few planets are very unusual and deserve special attention.

\paragraph{TOI 1690 (WD 1856+534)}  TOI 1690 b is a giant planet candidate transiting a white dwarf star \citep{wd1856}. TOI 1690 b is a landmark discovery because its existence demonstrates that planets can survive migration into  orbits close to white dwarf stars. With an orbital period of 1.4 days, TOI 1690 b ($R_p = 10.4 \pm 1 \rearth$, $M_p \lesssim 14 M_{Jupiter}$)  is close enough to its white dwarf host star that it must have formed at larger planet-star separations in order to avoid destruction when the white dwarf precursor star evolved into a red giant star.

\paragraph{TOI 849} TOI 849 b is a dense and unusually massive planet ($40.8_{-2.5}^{+2.4} \mearth$) for its size (smaller than Neptune, $3.447 _{-0.122}^{+0.164} \rearth$), such that it is considered a planetary core of a hot Jupiter \citep{armstrong20}. Of further interest is that TOI 849 b resides in the hot-Neptune ``desert,'' a region in exoplanet mass-radius period parameter space that contains few planets. TOI 849 b transits a G-type star  with an orbital period of 18.4 hours.

\paragraph{TOI 1266} The TOI 1266 system consists of two planets orbiting a nearby M-dwarf star \citep{stefansson20, demory20}. TOI 1266 is a key find because it will provide a strong test of photoevaporation models, as the outer planet ($P = 18.8$ days; $R_p = 1.67_{-0.11}^{+0.09} \rearth$) is very atypically smaller than the inner planet ($P = 10.9$ days, $R_p = 2.46 \pm 0.08 \rearth $). The outer and inner planets have upper mass limits of $6.4 \mearth$ and $15.9 \mearth$ respectively.

\section{Summary}
\label{sec-summary}

This paper described the process to identify \tess\ Objects of Interest, targets the project deems as promising planet candidates. The TOI process has yielded \tois\ TOIs over the course of the \tess\ prime mission. We gave an overview of the two major pipelines contributing targets for consideration as TOIs, the SPOC and QLP. The SPOC pipeline uniquely detected 1255 TOIs in the postage-stamp data and the QLP pipeline uniquely detected 986 TOIs in the FFI-only data. 

We illustrated the distribution of planets and planet candidates amassed over the \tess\ Prime Mission into the TOI Catalog and highlighted several \tess\ planets.
Ongoing follow-up efforts are adding to the growing collection of confirmed \tess\ planets As of \accessed, 72 TOIs have been confirmed and 1343 TOIs remain as active planet candidates.

Simulations of the first extended mission yield of planets predict the total number of planet candidates will double again. The two-year extended mission will run from July 2020 to September 2022. In the first year, \tess\ will re-observe the Southern Hemisphere in 13 sectors filling in the gaps from the Prime Mission. In the second year, \tess\ will observe 11 Northern Hemisphere sectors and 5 sectors on the ecliptic, covering two-thirds of the ecliptic plane. Additionally, the \tess\ ecliptic observations will overlap with 15 of the 20 K2 campaigns. The extended mission will increase the total sky coverage of the \tess\ mission overall from $\sim 70\%$ to $88\%$. With the addition of extended mission data, the TOI Catalog will gain additional TOIs and CTOIs, as well as improved planetary parameters for known short- and long-period TOIs. 

The TOI Catalog is posted online at the \url{tess.mit.edu/toi-releases} TOI portal. The TOI Catalog is a living list that is continually updated. An augmented version of the TOI catalog (updated twice daily) can be found at ExoFOP-TESS. MAST also  hosts archived versions of the catalog (saved monthly). The TOI Catalog will support the enduring legacy of the \tess\ mission by providing the list of planet systems which the exoplanet investigations of the next few decades will target.

\section*{Acknowledgements}

We thank the referee for their thoughtful comments which greatly improved the clarity of the paper.

Funding for the \tess\ mission is provided by NASA's Science Mission directorate. 

We thank William Fong and Martin Owens for their contributions to the TSO software.

This research has made use of the \tess\ Exoplanet Follow-up Observation Program website (ExoFOP-TESS), which is operated by the California Institute of Technology, under contract with the National Aeronautics and Space Administration under the Exoplanet Exploration Program.

This research has made use of the NASA Exoplanet Archive, which is operated by the California Institute of Technology, under contract with the National Aeronautics and Space Administration under the Exoplanet Exploration Program.

This paper includes data collected with the \tess\ mission, obtained from the MAST data archive at the Space Telescope Science Institute (STScI).
The analysis for this paper was done in part on the TESS Science Platform, a JupyterHub environment on Amazon Web Services that was deployed by the Space Telescope Science Institute (STScI). STScI is operated by the Association of Universities for Research in Astronomy, Inc., under NASA contract NAS 5–26555.


This work has made use of data from the European Space Agency (ESA) mission {\it Gaia} (\url{https://www.cosmos.esa.int/gaia}), processed by the {\it Gaia} Data Processing and Analysis Consortium (DPAC, \url{https://www.cosmos.esa.int/web/gaia/dpac/consortium}). Funding for the DPAC has been provided by national institutions, in particular the institutions participating in the {\it Gaia} Multilateral Agreement. 


This research has made use of NASA’s Astrophysics Data System.

Resources supporting this work were provided by the NASA High-End Computing (HEC) Program through the NASA Advanced Supercomputing (NAS) Division at Ames Research Center for the production of the SPOC data products.


A portion of this research was carried out at the Jet Propulsion Laboratory, California Institute of Technology, under a contract with the National Aeronautics and Space Administration (80NM0018D0004)


AV's work was performed under contract with the California Institute of Technology / Jet Propulsion Laboratory funded by NASA through the Sagan Fellowship Program executed by the NASA Exoplanet Science Institute.

D. D.'s work was supported in part by TESS Guest Investigator Program grant 80NSSC19K1727, and by NASA through Hubble Fellowship grant HST-HF2-51372.001-A awarded by the Space Telescope Science Institute, which is operated by the Association of Universities for Research in Astronomy, Inc., for NASA, under contract NAS5-26555.

HK acknowledges funding for the Stellar Astrophysics Centre provided by The Danish National Research Foundation (Grant agreement no.: DNRF106)

I.J.M.C. acknowledges support from the NSF through grant AST-1824644, and from NASA through Caltech/JPL grant RSA-1610091.
This work is partly supported by JSPS KAKENHI Grant Numbers JP18H01265 and JP18H05439, and JST PRESTO Grant Number JPMJPR1775.

KS acknowledges support from NASA 17-XRP17 2-0024.

CXH and MNG acknowledges support from MIT's Kavli Institute as a Juan Carlos Torres Fellow.

TD acknowledges support from MIT's Kavli Institute for Astrophysics and Space Research as a Kavli postdoctoral fellow.

We acknowledge Indigenous Peoples as the traditional stewards of the land, and the enduring relationship that exists between them and their traditional territories. The land on which this research was conducted is the traditional unceded territory of the Wampanoag Nation. We acknowledge the painful history of genocide and forced occupation of their territory, and we honor and respect the many diverse indigenous people connected to this land on which we gather from time immemorial.

\facility{TESS, MAST, NASA Exoplanet Archive, Gaia}

\software{QLP \citep{huanghlsp1, huanghlsp2}, 
    SPOC, \citep{jenkins:KDPH_TPS, Twicken2016},
    tica (Fausnaugh et al. (in prep.)),
    FITSH \citep{fitsh},
    astronet \citep{yu2019},
    TEC,
    TEV,
    batman \citep{batman},
    nebuliser \citep{Irwin:1985},
    emcee \citep{emcee},
    vartools \citep{vartools},
    go \citep{meyerson2014go},
    TensorFlow \citep{tensorflow2015-whitepaper},
    h5py \citep{collette_python_hdf5_2014},
    Astropy \citep{astropy:2013, astropy:2018},
    astroquery \citep{astroquery},
    matplotlib \citep{matplotlib},
    pandas \citep{reback2020pandas},
    Scipy \citep{2020SciPy-NMeth},
    Numpy \citep{numpy}
    }

\appendix
\section{List of Acronyms}
\begin{itemize}
    \item BLS, Box Least Squares (algorithm)
    \item BTJD, Barycentric-corrected \tess\ Julian Date 
    \item CAL, Calibration (module)
    \item CDPP, Combined Differential Photometric Precision
    \item CNN, Convolutional Neural Network
    \item COA, Compute Optimal Apertures (module)
    \item CP, Confirmed Planet (disposition)
    \item CTL, Candidate Target List
    \item CTOI, Community TOI
    \item DSN, Deep Space Network
    \item DV, Data Validation
    \item EB, Eclipsing Binary (disposition)
    \item ExoFOP, Exoplanet Follow-up Observing Program
    \item FFI, Full-Frame Image
    \item FP, False Positive (disposition)
    \item IS, Instrument Noise/Systematic (disposition)
    \item \jwst, James Webb Space Telescope
    \item KP, Known Planet (disposition)
    \item MAP, Maximum A Posteriori (fit)
    \item MAST, Mikulksi Archive for Space Telescopes
    \item MES, Multiple Event Statistic
    \item PA, Photometric Analysis (module)
    \item PC, Planet Candidate (disposition)
    \item PDC, Presearch Data Conditioning (module)
    \item PDCSAP, PDC Simple Aperture Photometry (light curve)
    \item POC, Payload Operations Center
    \item PPA, Photometer Performance Assessment
    \item PRF, Pixel Response Function
    \item PRV, precise Radial Velocity
    \item QLP, Quick Look Pipeline
    \item RV, Radial Velocity
    \item SES, Single Event Statistic
    \item SNR, Signal-to-Noise Ratio
    \item SOC, Science Operations Center
    \item SPOC, Science Processing Operations Center 
    \item SVD, Singular Value Decomposition
    \item TCE, Threshold Crossing Event
    \item TEC, \tess-ExoClass
    \item TEV, \tess~Exoplanet Vetter
    \item \tess, Transiting Exoplanet Survey Satellite
    \item TSM, Transmission Spectroscopy Metric
    \item TFOP, \tess~Follow-Up Observing Program
    \item TIC, \tess~Input Catalog
    \item TJD, \tess~Julian Date
    \item TOI, \tess~Object of Interest 
    \item TPS, Transiting Planet Search (module)
    \item TSO, \tess~Science Office
    \item U, Undecided (disposition)
    \item V, Stellar Variability (disposition)
\end{itemize}

\bibliographystyle{apj}
\bibliography{refs}

\end{document}